\documentstyle[11pt,epsfig]{article}
\def\thefootnote{\fnsymbol{footnote}}
\def\bm#1{\mbox{\boldmath$#1$\/}}

\def\NNL{{N${}^2$L}}
\def\roughly#1{\mathrel{\raise.3ex\hbox{$#1$\kern-.75em%
\lower1ex\hbox{$\sim$}}}}
\def\lsim{\roughly<}

\def\fpi{f_\pi}
\def\mpi{m_\pi}
\def\TminusS{{{\hat T}_S^{(-)}}}
\def\TminusT{{{\hat T}_T^{(-)}}}
\def\TplusS{{{\hat T}_S^{(+)}}}
\def\TplusT{{{\hat T}_T^{(+)}}}
\def\Tr{{\mbox{Tr}}}
\def\TtimesS{{{\hat T}_S^{(\times)}}}
\def\TtimesT{{{\hat T}_T^{(\times)}}}
\def\ddm{\delta m_{\!\Delta}}
\def\mN{m_{{}_{\rm N}}}
\def\Jem{J_{\!\rm em}}

\def\ttz{(\tau_1\times\tau_2)^z}
\def\tt{\tau_1\times\tau_2}
\def\muT{\mu_{\rm T}}
\def\BplusS{B_{\rm S}^{(+)}}
\def\BplusT{B_{\rm T}^{(+)}}
\def\gL{g_{\mbox{\tiny L}}}
\def\gR{g_{\mbox{\tiny R}}}
\def\ldiv{\eta}
\def\qdiv{\Delta(m_\pi^2)}
\def\bea {\begin{eqnarray}}
\def\eea {\end{eqnarray}}
\def\be {\begin{equation}}
\def\ee {\end{equation}}
\def\bitem{\begin{itemize}}
\def\eitem{\end{itemize}}
\def\ie{{\it i.e}}
\def\del{\partial}
\def\M{{\cal M}}
\def\O{{\cal O}}

\def\calP{{\cal P}}
\def\calD{{\cal D}}
\def\L{{\cal L}}
\def\gA{g_A}

\def\Nbar{{\overline {N}}}
\def\Bbar{{\overline {B}}}
\def\bfK{{\bf K}}
\def\dslash{{\not {\! \partial}}}
\def\vslash{{\not {\! v}}}
\def\tauvec{{\vec \tau}}
\def\dmu{{\partial_\mu}}
\def\GpiV{{\Gamma_{\pi V}^{\mu,ab}}}
\def\GpiVct{{\Gamma_{\pi V,\mbox{\tiny CT}}^{\mu,ab}}}
\def\prl {Phys. Rev. Lett.}
\def\pl {Phys. Lett.}
\def\pr {Phys. Rev.}
\def\np {Nucl. Phys.}
\newcommand{\e}{{\mbox{e}}}
\newcommand{\zero}[1]{{\stackrel{\mbox{\tiny o}}{#1}}}

\begin{document}

\begin{titlepage}
\begin{center}
\begin{flushright}
SNUTP 95-043
\end{flushright}
\hfill\today
\vskip 1.0cm
{\large\bf Chiral Lagrangian Approach to}
\vskip 0.2cm
{\large\bf Exchange Vector Currents in Nuclei}
\vskip 2cm
{\large Tae-Sun Park$^{(a)}$, Dong-Pil Min$^{(a)}$ and Mannque Rho$^{(b)}$}\\
\vskip 0.2cm
{\it Institute for Nuclear Theory, University of Washington}\\
{\it Seattle, WA 98195, U.S.A.}
\vskip 2cm
{\bf ABSTRACT}
\begin{quotation}
\noindent
Exchange vector currents are calculated up to one-loop order (corresponding
to next-to-next-to-leading order) in chiral perturbation theory.
As an illustration of the power of the approach, we apply the formalism to
the classic nuclear process $n+p\rightarrow d +\gamma$ at thermal energy.
The exchange current correction comes out to be $(4.5 \pm 0.3)$ \% in
amplitude giving a predicted cross section $\sigma= (334\pm 3)\ {\mbox mb}$ in
excellent agreement with the experimental value $(334.2\pm 0.5)\ {\mbox mb}$.
Together with the axial charge transitions computed previously,
this result provides a strong support for
the power of chiral Lagrangians in nuclear physics. As a by-product of
our results, we suggest an open problem in the application of chiral
Lagrangian approach
to nuclear processes that has to do with giving
a physical meaning to the short-range correlations that
play an important role in nuclei.
\end{quotation}
\end{center}

\vfill
\noindent
$(a)$ Permanent address :
Department of Physics and Center for Theoretical Physics,
Seoul National University, Seoul 151-742, Korea

\noindent
$(b)$ Permanent address :
Service de Physique Th\'{e}orique, CEA Saclay,
91191 Gif-sur-Yvette, France

\end{titlepage}
\renewcommand{\thefootnote}{\#\arabic{footnote}}
\setcounter{footnote}{0}
\section{Introduction}
\indent

Nuclear responses to electroweak probes
are described by matrix elements of the corresponding currents.
The major contribution comes from one-body currents, specifically,
the sum of the currents of each nucleon.
The principal corrections come from
exchange currents (or meson-exchange currents) that consist
of two-body and higher-body currents.
By now there exist a large number of unambiguous
experimental evidence \cite{fm} for the presence and structure of
meson-exchange currents.
While phenomenological in character,
the approaches taken so far
can describe the large bulk of experiments \cite{review}
rather successfully,
providing us
our progressive understanding of nuclear processes.
To the extent that nucleon-nucleon interactions are now fairly accurately
understood, one can have a great deal of confidence in
the theoretical tool with which the effect of exchange currents
is calculated.
There remains however the fundamental question as to how
our phenomenological understanding of nuclear forces and associated
meson currents can be linked
to the fundamental theory of strong interactions, QCD.

In the modern understanding of QCD, it is the spontaneous
breaking of chiral symmetry associated with the light quarks that
predominantly governs the structure of low-energy hadrons
as well as the forces mediating between them. In fact, the full content
of the gauge theory of strong interactions, QCD, can be expressed
at low energy by a systematic chiral expansion starting with effective
chiral Lagrangians \cite{wein79}. Stated more strongly,
such an approach, known as chiral perturbation theory
(ChPT),\ while reproducing the current algebra,
is now considered to be {\it exactly} equivalent to QCD in long
wavelength regime \cite{leut94}.
The question that is immediately posed here is whether or not
one can describe nuclear processes from the chiral perturbation
theory  point of view.

Two recent developments suggest that the answer to this question is
affirmative.
The first is the work of Weinberg \cite{wein90} and Ord\'o\~nez, Ray and
van Kolck \cite{vankolck} on understanding nuclear forces from chiral
Lagrangians.
The second is the explanation by the present authors \cite{pmr1}
of the enhanced axial-charge
transitions in heavy nuclei in terms of
exchange axial currents in chiral perturbation theory
treated to the same chiral order as for
nuclear forces.
Both of these two works provide a systematic
and consistent field-theory-based understanding of the nuclear processes
and validate at the same time the traditional approaches to the processes.
In this paper, we present one more case where the chiral Lagrangian
scores a stunning success, namely, the radiative neutron capture
\be
n+p\rightarrow d+\gamma
\label{np} \ee
at threshold. This result was briefly reported in a Letter \cite{pmrnp}.
Here we shall go into greater details of the calculation involved for
the process (\ref{np}).

Historically, the process (\ref{np}) was first explained quantitatively
two decades ago by Riska and Brown \cite{riskabrown} who showed
that the $\sim 10 \%$ discrepancy
between the experimental cross-section and the theoretical impulse
approximation prediction is eliminated by exchange currents.
Riska and Brown computed, using realistic hard-core wavefunctions,
the two one-pion-exchange
diagrams initially suggested in 1947 by Villars \cite{villars}
plus the $\omega$ and $\Delta$ resonance diagrams. That the dominant
contributions to electroweak exchange currents could be gotten from
current-algebra low-energy theorems was suggested by
Chemtob and Rho \cite{chemrho} who gave a systematic rule for
organizing the leading exchange-current diagrams effective
at low energy and momentum.
Although suspected since the Yukawa force was introduced, the work of
Riska and Brown was the first unequivocal evidence for the role of mesons,
in particular that of pions, in nuclear interactions.
In this paper, we show that the terms considered by Riska and Brown
are a main part of the terms that figure
in chiral perturbation theory to
next-to-next-to-leading (\NNL) order
and that when completed by the rest of the \NNL\ order
terms, chiral perturbation theory works impressively,
confirming on the one hand the work of Riska and Brown \cite{riskabrown}
and on the other hand the earlier (``chiral filter") conjecture
of Kubodera, Delorme and Rho \cite{kdr}.

In all cases studied so far, one is limited to long wavelength
processes, with the typical energy/momentum scale $Q$ much less than
the chiral symmetry scale
$\Lambda_\chi\sim 4\pi f_\pi \sim 1$ GeV. This is because
ChPT is an expansion in $Q/\Lambda_\chi$ and its practical value lies
where $Q/\Lambda_\chi\ll 1$.
For nuclear physics, where both baryons and mesons enter on the same footing,
the heavy-fermion formalism as discussed in \cite{wein90,HFF} proves to be
convenient for making a systematic expansion in derivative on pion fields
as well as on baryons fields, $\del / \Lambda_\chi$ and in
$m_\pi / \Lambda_\chi$ where $m_\pi$ is the pseudo-Goldstone boson mass.
In this formalism, the nucleon mass is regarded as heavy,
$\mN \simeq \Lambda_\chi$.\footnote{In
principle, one can formulate chiral perturbation theory equally well
in relativistic form
for the baryons provided that care is exercised in arranging terms in
consistency with chiral counting. At low nuclear matter density, however, the
relativistic formulation is awkward and the heavy-fermion formalism
is a lot more powerful and convenient in implementing in nuclear problems of
the sort that we are concerned with. The situation will be different
at higher matter density. This is because at higher matter density at
which the nucleon mass is effectively reduced as discussed {\it e.g.}, by
Brown and Rho \cite{br91,br95}, the heavy-baryon strategy may be
suspect as the ``$1/m_N$" expansion may not converge rapidly enough.
One of the authors (MR) is grateful to Heiri Leutwyler for a discussion on
this matter.}

As in \cite{wein90},
the expansion is organized by the power $\nu$ in $Q^\nu$.
This expansion will be reviewed in the next section.
Here we remark briefly how kinematical considerations affect the
chiral counting rule, modifying it from the naive counting rule.
The modification comes from the fact that the space part of the vector
current and the time part of the axial-vector current
$(V^i,\, A^0)$ are proportional to
$(\gamma^i,\,\gamma_5\gamma^0)$ and hence are
suppressed compared to the other part of
the currents, $(V^0,\,A^i)$.
The correct counting for one-body currents
in two-body systems is $(V^i,\,A^0)_{\rm 1B}= \O(Q^{-2})$.
Compared to these one-body currents,
the leading two-body current contributions
are of order $\O(Q)$
and the one-loop corrections are of order $\O(Q^3)$.
Thus from the point of view of the chiral expansion,
the two-body currents at one-loop order
correspond to next-to-next-to-leading (\NNL) order.
This is the order that was computed in the case of the nuclear
axial-charge transitions studied in \cite{pmr1}
and that will be adopted for the vector current matrix element for the
process (\ref{np}).
Three-body (and higher-body) currents are suppressed for
exactly the same reason as for the suppression of
3-body forces discussed by Weinberg \cite{wein90}:
the resulting 3-body currents are of order
$\O(Q^4)$ compared to the 1-body currents.

Given these counting rules,
the ChPT in nuclear systems can be made to correspond to computing
Feynman diagrams involving external fields
in the increasing power $\nu$ {\it embedded} inside the most general
process describing the transition from the initial nuclear state to the
final nuclear state with interactions taking place before and after the
current insertion. This strategy is essentially the same as
what was first suggested, with somewhat
{\it ad hoc} assumptions, in \cite{chemrho}. What this means in practice is
that {\it we are to take the most realistic nuclear wave functions and
calculate the transition matrix elements with the ChPT graphs computed
to the maximum possible order of chiral expansion.} This allows us
to fix the ``counter terms" in the chiral expansion from experiments.
At the present stage of our
understanding on the role of chiral symmetry in
nuclear systems, this seems to be the only practical attitude to take
in confronting nuclear dynamics. We might point out that this point of
view is consistent with the approach advocated
by Weinberg \cite{wein90} in his treatment of nuclear
forces, in particular many-body forces.
An important consequence of this strategy is then that only the current
operators obtainable by ChPT are to be kept.
This implies mainly two things.
Firstly,
it implies that short-wavelength effects are to be ``filtered out".
Secondly, it implies that
$n$-body currents with $n>2$ are suppressed in nuclear systems with
the mass number $A>2$ in the same sense
that $n$-body forces are suppressed \cite{wein90}.
Of particular importance of the first implication is that when the
currents are put in coordinate space, shorter-range interactions for
$r_{12} \lsim r_c$ -- where $r_c$ is the hard core radius --
cannot contribute in
ChPT. This is because the ``short-range correlations" that play an important
role in nuclear observables represent those degrees of freedom that
cannot be accessed by chiral perturbation expansion.
Clearly this strategy would make sense only if the result does not sensitively
depend on the precise value of the ``hard-core" size $r_c$, within the
relevant range for
application in nuclei to the order considered, say,
$\Lambda_\chi^{-1} \lsim r_c \lsim (2m_\pi)^{-1}$.
Note that this is roughly the range that can be reasonably described in
ChPT for the NN potential \cite{vankolck}.

This paper is organized as follows. In section 2, chiral perturbation theory
as appropriate for the problem at hand will be briefly summarized.
The main purpose of this section is to define the notations and indicate
the strategy specific to nuclear problems as outlined above and detailed
in \cite{pmr1}. For a general and
comprehensive review of chiral perturbation theory, we refer to recent
review articles \cite{reviewchp}. Section 3 deals with the derivation of
the exchange electromagnetic currents in nuclei to {\NNL} order for
vanishingly small momentum transfers. The formulas derived in Section 3
are applied in Section 4 to the classic nuclear process (\ref{np}).
Some further remarks and conclusion are given in Section 5. The Appendices
give detailed formulas used in the main text.

\section{Chiral Perturbation Theory}
\indent

Here we shall briefly review the specific aspect of
chiral perturbation theory (ChPT) adopted for our calculation to one-loop
order.
This section is organized as follows.
We start, in Section~\ref{counting:S}, with a rederivation of Weinberg's
counting rules with a slight generalization. The relevant
chiral Lagrangians in heavy-fermion formalism (HFF) will be written down
in Section~\ref{HFF:S} and the renormalization procedure
within HFF in Section~\ref{renormalization:S}. The particular feature
of ChPT in nuclear physics will be described in Section~\ref{nuclear:S}.

\subsection{Counting rules}
\label{counting:S}
\indent

Here we rederive and generalize somewhat Weinberg's counting
rule \cite{wein90} along the line described in \cite{pmr1}.
Although we shall not consider explicitly the
vector-meson degrees of freedom,
we include them here in addition to pions and nucleons.
Much of what we will obtain later without vector mesons
turn out to be valid in their presence. In our work, we consider
the vector-meson masses -- which are comparable to the
chiral scale $\Lambda_{\chi}$ --  as
heavy compared to the momentum probed $Q$ -- say, scale of external
three momenta or $m_\pi$. In dense medium, the situation is different
as discussed in \cite{br95}. There the vector-meson masses
can be considered small, in the sense of Georgi's vector limit.
In our discussions, we do not encounter the dense regime in which
vector-meson masses drop as a function of density.

In establishing the counting rule, we make the following assumptions:
Every intermediate meson (whether heavy or light) carries a four-momentum
of order of $Q$. In addition we assume that
for any loop, the effective cut-off in the loop integration
is of order of $Q$. We will specify what this means physically
when we discuss specific processes, for this clarifies the limitation of
the chiral expansion scheme.

An arbitrary Feynman graph can be characterized by the number $E_N (E_H, E_E)$
of external -- both incoming and outgoing -- nucleon
(vector-meson, external field) lines, the number $L$ of loops, the number
$I_N (I_\pi,\ I_H)$ of internal nucleon (pion, vector-meson) lines
and the number of $C$ disconnected pieces of the Feynman graphs.
For an connected graph, for example, we should have $C=1$.
Each vertex can in turn be  characterized by
the number $d_i$ of derivatives and/or of $m_\pi$ factors and
the number $n_i$ $(h_i, e_i)$ of nucleon
(vector-meson, external field) lines attached
to the vertex.
Now for a nucleon intermediate state of  momentum $p^\mu=\mN v^\mu + k^\mu$
where $k^\mu = {\cal O}(Q)$ and $v^\mu$ is a
constant 4-vector with $v^\mu\simeq (1,\,{\bm 0})$,
we acquire a factor $Q^{-1}$ since
\be
S_F(\mN v+k) = \frac{1}{v\cdot k} = {\cal O}(Q^{-1}),
\ee
where we have used the propagator obtained by the HFF which will be explained
in the next section.
An internal pion line contributes a factor $Q^{-2}$ since
\be
\Delta_F(q^2;m_\pi^2)=\frac{1}{q^2 - m_\pi^2} = {\cal O}(Q^{-2})
\nonumber
\ee
while a vector-meson intermediate state contributes
$Q^0$ $(\sim O(1))$ as one can see from its propagator,
noting that
$D_F^{\mu\nu} (q^2;m_V^2) =
\left(-g^{\mu\nu} + \frac{q^\mu q^\nu}{m_V^2}\right)
\Delta_F(q^2; m_V^2)
= \O(Q^0) \,\Delta_F(q^2; m_V^2)$,
\be
\Delta_F (q^2;m_V^2) = \frac{1}{q^2 - m_V^2}
\simeq \frac{1}{-m_V^2} = {\cal O}(Q^0)
\ee
where $m_V$ represents a generic mass of vector mesons.
And a loop contributes a factor $Q^4$ because
its effective cut-off is assumed to be of order of $Q$.
Finally, the $C$ separated pieces of the Feynman graph
contributes a factor $Q^{4(1-C)}$ due to the $(C-1)$ times of the
energy-momentum conservation factor, $(2\pi)^4 \delta^4(\sum_i p_i)$
where $(\sum_i p_i)$ denote the sum of
incoming and outgoing momenta of a separate piece of
the Feynman graph.

We thus arrive at the counting rule that an arbitrary graph is characterized
by the factor $Q^\nu$ with
\be
\nu = - I_N - 2 I_\pi + 4 L + 4 - 4 C +\sum_i d_i
\label{piN0}\ee
where the sum over $i$  runs over all vertices of the graph.
Using the identities,
$I_\pi + I_H + I_N = L + V -C$
where $V$ denotes the total number of vertices of the
Feynman graph,
$I_H = \frac12 (\sum_i h_i - E_H)$,
$I_N = \frac12 (\sum_i n_i - E_N)$
and $\sum_i e_i = E_E$,
we can rewrite the counting rule
\be
\nu = 4 - 2 C  - \left(\frac{E_N}{2}  +  E_H + E_E \right) + 2 L
 + \sum_i {\bar \nu}_i,\ \ \ \ \
{\bar \nu}_i \equiv d_i + \frac{n_i}{2} + h_i  + e_i - 2 .
\label{counting}\ee
We recover the counting rule derived by Weinberg \cite{wein90} if we set
$E_H=h_i=0$ and $E_E= e_i=0$.
With non-zero external heavy meson lines,
it is generally hard to satisfy our condition that
all the internal lines should carry momenta of order of $Q$.
But there is no limitation on the number of $h_i,\,e_i$ and $E_E$.

The quantity ${\bar \nu}_i$ is defined so that
\be
{\bar \nu}_i =\left(  d_i + \frac{n_i}{2} + h_i  + e_i - 2 \right) \ge 0.
\label{barnu}\ee
This is guaranteed by chiral symmetry \cite{wein90}
even in the presence of external fields.
In the absence of external fields (as in nuclear forces),
\be
d_i + \frac{n_i}{2} + h_i - 2 \geq 0.
\ee
This means that the
leading order effect comes from graphs with vertices satisfying
\be
d_i + \frac{n_i}{2} + h_i - 2 = 0\,.
\ee
Examples of vertices of this kind are:
$\pi^k NN$ with $k\geq 1\, (d_i=1,\ n_i=2,\ h_i=0)$,
$h N N\, (d_i=0,\ n_i=2,\ h_i=1)$,
four-Fermi contact interactions,
$(d_i=0,\ n_i=4,\ h_i=0)$,
$h\pi^k$ with $k\geq 1 \,(d_i=1,\ n_i=0,\ h_i=1)$, etc where $h$ denotes
vector-meson fields.
In $NN$ scattering or in nuclear forces,
$\frac{E_N}{2} =2$ and $E_H=0$, and so we have
$\nu \geq 0$.
The leading order contribution corresponds to $\nu=0$,
coming from three classes of diagrams; one-pion-exchange,
one-vector-meson-exchange and four-Fermi contact graphs.
In $\pi N$ scattering, $\frac{E_N}{2} =1$ and $E_H=0$, we have
$\nu \geq 1 $ and the  leading order comes from
nucleon Born graphs, seagull graphs and one-vector-meson-exchange
graphs.

In the presence of external fields denoted generically ${\cal E}$,
the condition becomes \cite{mr91}
\be
\left( d_i + \frac{n_i}{2}  + h_i-2\right) \geq -1 \,.\label{exchcond}
\ee
The difference from the previous case comes from the fact that a derivative
is replaced by a gauge field.
The equality holds only for the vertices
${\cal E} \pi^j NN$ with $j \geq 0$ ($d_i=h_i=0,\,n_i=2$),
${\cal E} h \pi^j$ with $j\geq 0$ ($h_i=1,\,n_i=d_i=0$)
and ${\cal E} \pi^k$ with $k \geq 1$ ($h_i=n_i=0$, $d_i=1$).
We will later show that this is related to the ``chiral filter" phenomenon.
The condition (\ref{exchcond}) plays an important role in determining
exchange currents.
Apart from the usual nucleon Born terms which are in the class of
``reducible" graphs and hence do not enter into our consideration,
we have two graphs that contribute in the leading order to the exchange
current: the ``seagull" graphs and ``pion-pole" graphs \footnote{These are
standard jargons in the literature. See \cite{review,chemrho}.},
both of which involve a vertex with ${\bar \nu}_i=0$.
On the other hand, a vector-meson-exchange graph
involves a ${\bar \nu}_i= 2$ vertex. This is because $d_i=1,\ h_i = 2$
at the ${\cal E} hh$ vertex and
$d_i=1$, $n_i=2$, $h_i=1$ at the ${\cal E} h NN$ vertex,
where ${\cal E}$ denotes external fields.
Therefore vector-exchange graphs are suppressed
by power of $Q^2$. This counting rule is the basis for establishing
the chiral filtering even when vector mesons are present
(see Appendix~\ref{Res:A}).
Thus the results we obtain without explicit vector mesons are valid
more generally.

We now focus on the nuclear responses to electroweak probes.
Denoting the vector ($V^\mu$) and axial-vector currents ($A^\mu$)
collectively by $G^\mu$, we decompose the latter in terms of the number of
nucleons involved:
\be
G^\mu = G^\mu_{\rm 1B} + G^\mu_{\rm 2B} + \cdots.
\ee
In principle, there will be $n$-body currents for $N \ge n \ge 1$
in $N$-body systems.
Compared to the leading 2-body currents,
the leading $3$-body currents are suppressed by
a factor $\O(Q^2/\Lambda_\chi^2)$ from
the counting rule eq.(\ref{counting})
and  further suppressed by a factor
$\O(Q/\Lambda_\chi)$ for
exactly the same reason as for the
suppression of the 3-body forces \cite{wein90}.
Thus the net suppression factor is of order
$(Q/\Lambda_\chi)^3$ compared to the 2-body currents. We can therefore safely
ignore three-body and higher-body currents.
Limiting ourselves to the two-body subsystem, we have
$E_H=0$, $\frac{E_N}{2}=2$ and $E_E=1$. The
counting rule reads
\be
\nu = 1 - 2 C  + 2 L
 + \sum_i {\bar \nu}_i,\ \ \ \ \
{\bar \nu}_i \equiv d_i + \frac{n_i}{2} + h_i  + e_i - 2
\label{currentcounting}\ee
where $C=2$ for one-body currents and $C=1$ for two-body currents.
This counting rule would naively imply
$G^\mu_{\rm 1B}= \O(Q^{-3})$ and $G^\mu_{\rm 2B}= \O(Q^{-1})$.
But as mentioned in Introduction, there is a further suppression factor
of order $Q/\Lambda_\chi$ for
$(V^i,\, A^0)_{\rm 1B}$ and $(V^0,\, A^i)_{\rm 2B}$,
with $i=1,\,2,\,3$.
The suppression of one-body currents can be understood
simply by recalling that
the leading-order vector and axial-vector one-body currents are
proportional, respectively, to
${\bar u} \gamma^\mu u = (\O(1), \O(Q/\mN) )$ and
${\bar u} g_A \gamma^\mu \gamma_5 u = (\O(Q/\mN), \O(1))$
where $u$ denotes the spinor of the nucleon.
As for the two-body currents, the same suppression factor appears
with the interchange of the role of
the vector and axial-vector currents.
The effective chiral counting for the currents is summarized
in Table~\ref{count:T}.
\begin{table}[t]
\begin{center}
\begin{tabular}{|c||c|c||c|}\hline
  & $V^i$ and $A^0$ & $V^0$ and $A^i$ & $G^\mu$ ({\it naive})\\
\hline \hline
1-body & $\O(Q^{-2})$ & $\O(Q^{-3})$ & $\O(Q^{-3})$ \\ \hline
2-body (leading) & $\O(Q^{-1})$ & $\O(Q^{0})$ & $\O(Q^{-1})$\\ \hline
2-body (1-loop correction) & $\O(Q^1)$ & $\O(Q^1)$ & $\O(Q^{1})$ \\ \hline
\hline
$\langle M_{\rm 2B} \rangle / \langle M_{\rm 1B} \rangle$
 & $\O(Q^1)$ & $\O(Q^3)$ & $\O(Q^2)$ \\ \hline
\end{tabular}
\caption{The chiral counting for the vector and axial-vector currents
for two-body subsystems.
The last row gives the ratio of the amplitude
of the two-body currents compared to that of the one-body
currents. The last column shows the naive counting. }
\label{count:T}
\end{center}
\end{table}
One can see immediately that
the effect of the two-body currents can be sizable
while the loop corrections are small for the space component
of the vector current and the time component of
the axial-vector current. This feature was noted sometime
ago \cite{kdr}.

\subsection{Effective chiral Lagrangian in heavy-baryon formalism}
\label{HFF:S}
\indent

The effective chiral Lagrangian that consists of pions and nucleons
involving lowest derivative terms takes the form \cite{gss},
\bea
{\cal L} &=&
\Nbar\left[ i\gamma^\mu D_\mu - \mN
+ i \gA \gamma^\mu\gamma_5 \Delta_\mu\right] N
 - \frac{1}{2} \sum_A C_A \left(\Nbar \Gamma_A N\right)^2
\nonumber \\
&&+\, \frac{f_\pi^2}{4} {\rm Tr}\left(\nabla_\mu \Sigma^\dagger
\nabla^\mu \Sigma\right)
+ \frac{f_\pi^2}{4} m_\pi^2 {\rm Tr}(\Sigma + \Sigma^\dagger)
+ \cdots ,
\label{chiralag}
\eea
where
\bea
\Delta_\mu &=&
\frac12 \left\{ \xi^\dagger , \dmu \xi\right\}
- \frac{i}{2} \xi^\dagger ({\cal V}_\mu+{\cal A}_\mu) \xi
+ \frac{i}{2} \xi ({\cal V}_\mu - {\cal A}_\mu)\xi^\dagger,\nonumber\\
\nabla_\mu \Sigma &=& \dmu \Sigma - i ({\cal V}_\mu+{\cal A}_\mu) \Sigma
+ i \Sigma ({\cal V}_\mu - {\cal A}_\mu)
\eea
with the external gauge fields
${\cal V}_\mu= {\vec {\cal V}}_\mu \cdot \frac{\tauvec}{2}$ and
${\cal A}_\mu= {\vec {\cal A}}_\mu\cdot \frac{\tauvec}{2}$.
Here $\mN\simeq 939$~MeV is the nucleon mass, $\gA\simeq 1.25$
is the axial coupling constant, $f_\pi \simeq 93$~MeV is the pion decay
constant and $m_\pi \simeq 139$~MeV
is the pion mass.
The ellipsis stands for higher derivative and/or symmetry-breaking
terms that will be specified later as needed.
Under chiral SU(2)$\times$SU(2) transformation,
the chiral field $\Sigma= {\rm exp}(i\frac{\tauvec\cdot {\vec \pi}}{f_\pi})$
transforms as $\Sigma\rightarrow \gR \Sigma \gL^\dagger$
($\gR,\gL \in$ SU(2))
and the covariant derivative transforms the same as $\Sigma$
does,
\bea
\nabla_\mu \Sigma &=& \dmu \Sigma - i ({\cal V}_\mu+{\cal A}_\mu) \Sigma
+ i \Sigma ({\cal V}_\mu - {\cal A}_\mu)
\nonumber\\
&\rightarrow & \gR\, \nabla_\mu \Sigma \, \gL^\dagger
\eea
with the external gauge fields
transforming locally
\bea
{\cal V}_\mu+{\cal A}_\mu &\rightarrow&
{\cal V}_\mu' + {\cal A}_\mu' = \gR ({\cal V}_\mu + {\cal A}_\mu)\gR^\dagger
-i\dmu \gR \cdot \gR^\dagger,
\nonumber \\
{\cal V}_\mu-{\cal A}_\mu &\rightarrow&
{\cal V}_\mu' - {\cal A}_\mu' = \gL ({\cal V}_\mu - {\cal A}_\mu)\gL^\dagger
-i\dmu \gL \cdot \gL^\dagger.
\nonumber \label{vaLR}\eea
The non-linear realization of chiral symmetry is expressed in terms of
$\xi = \sqrt{\Sigma} = {\rm exp}(i\frac{{\vec \tau}\cdot {\vec \pi}}{2 f_\pi})$
and $h = h (\xi,\gL,\gR)$ defined with $\xi$
$$\xi \rightarrow \gR \xi h^\dagger = h \xi \gL^\dagger.$$
The nucleon field $N$ transforms as a matter field, {\it i.e},
$N\rightarrow h N$,
and its covariant derivative as
$D_\mu N \rightarrow h D_\mu N$ while
$\Delta_\mu \rightarrow h \Delta_\mu h^\dagger$ where\footnote{
We have defined two covariant derivatives involving chiral fields,
$\nabla_\mu \Sigma$ and $\Delta_\mu$.
The first transforms as $\Sigma$ does while the latter as
$N$ does.
We can express one in terms of the other,
$$
\Delta_\mu= \frac12 \xi^\dagger \left(\nabla_\mu \Sigma\right) \xi^\dagger,
\ \ \ \ \
\nabla_\mu \Sigma = 2 \xi \Delta_\mu \xi.
$$
We use, as is done frequently in the literature,
$\nabla_\mu \Sigma$ for the meson sector and $\Delta_\mu$ for the
meson-nucleon sector.}
\bea
D_\mu N &=& (\dmu + \Gamma_\mu) N ,
\nonumber \\
\Gamma_\mu &=& \frac{1}{2} \left[\xi^\dagger,\, \dmu \xi\right]
-\frac{i}{2}\xi^\dagger ({\cal V}_\mu+{\cal A}_\mu) \xi - \frac{i}{2}
\xi ({\cal V}_\mu - {\cal A}_\mu)\xi^\dagger.
\label{deltamu}\eea
Note that $h(\xi, \gL, \gR)$ is a complicated local function
of $\xi$, $\gL$ and $\gR$, the explicit form of which
is not needed.

We have included the four-Fermi non-derivative contact term
studied by Weinberg \cite{wein90}. We shall
ignore possible four-Fermi
contact terms involving quark mass terms and
derivatives except for counter terms needed
later. They are not relevant
to the chiral order (in the sense defined precisely later) that we are working
with. The explicit chiral symmetry breaking is included minimally in the
form of the pion mass term.
Higher order symmetry breaking terms do not
play a role in our calculation.

For completeness -- and to define our notations -- we sketch here
the basic element of the heavy-fermion formalism
(HFF) \cite{georgi} applied to nuclear systems as developed by Jenkins
and Manohar \cite{jm91} wherein the nucleon is treated as a heavy fermion.
As stressed in Introduction, the relativistic formulation of ChPT works
well when only mesons are involved but it does not work when baryons are
involved since while space derivatives on baryons fields can be arranged
to appear on the
same footing as four-derivatives on pion fields, the time derivative on baryon
fields picks up a term of order of the chiral symmetry breaking scale and hence
cannot be used in the chiral counting. This problem is avoided in the
HFF. To set up the HFF, the fermion momentum is written as
\be
p^\mu = \mN v^\mu + k^\mu
\ee
where $v^\mu$ is the $4$-velocity with $v^2=1$, and $k^\mu$ is the small
residual momentum. (In the practical calculation that follows, we will choose
the heavy-fermion rest frame $v^\mu =(1, {\bm 0})$.)
{}From the conventional nucleon field $N(x)$,
we define heavy fermion field $B_v(x)$ for a given
four-velocity $v^\mu$,
by
\be
B_v(x) = \e^{i \mN \, v\cdot x} N(x),
\ee
which we divide into two parts,
\be
B_v= B_v^{(+)} + B_v^{(-)} \equiv
\frac{1+\vslash}{2} B_v + \frac{1-\vslash}{2} B_v.
\ee
As defined,
$B_v^{(+)}(B_v^{(-)})$ can be identified as positive
(negative) energy solution.
Since the free Lagrangian for the nucleon is
\be
\Nbar( i \dslash - \mN) N =
\Bbar_v^{(+)}\, i \dslash\, B_v^{(+)}
+ \Bbar_v^{(-)} (i \dslash - 2 \mN) B_v^{(-)},
\ee
$B_v^{(+)}$ can be viewed as the relevant low-energy degree of
freedom (with a zero effective mass),
while $B_v^{(-)}$ as the irrelevant degree of freedom
(with an effective mass of $2 \mN$).
By integrating out the $B_v^{(-)}$,
we get an effective Lagrangian involving only the $B_v^{(+)}$ and pion fields.
Since
\be
\vslash B_v^{(+)}= B_v^{(+)},
\ee
we can reduce all the $4\times 4$ gamma matrices to
$2\times 2$ Pauli matrices.
Define the spin operators $S_v^\mu$ and $S_v^{\mu\nu}$ by
\bea
S_v^\mu &\equiv& \frac14 \gamma_5\left[\vslash,\,\gamma^\mu\right].
\\
S_v^{\mu\nu} &\equiv& \left[S_v^\mu,\,S_v^\nu\right]
= i \epsilon^{\mu\nu\alpha\beta} v_\alpha S_{v\beta}.
\eea
In the rest frame, $v^\mu= (1,\, {\bm 0})$, so
$S_v^0 = S_v^{0i} = 0$ and
\bea
S_v^i &=& \frac12 \sigma^i,
\nonumber \\
S_v^{ij} &=& \frac{i}{2} \epsilon^{ijk} \sigma^k ,
\ \ \ \ \epsilon^{123}=1.
\eea
Denoting $B_v^{(+)}$ by $B$ and omitting the subscript $v$,
our chiral Lagrangian (\ref{chiralag}) expressed
in terms of the heavy-fermion field to leading ({\ie}, zeroth)
order in $\frac{1}{\mN}$ and in chiral order takes the form
\bea
{\cal L}_0 &=& \Bbar\left[ i v\cdot D
+ 2 i \gA S\cdot \Delta \right] B
- \frac{1}{2} \sum_A C_A \left(\Bbar \Gamma_A B\right)^2
\nonumber \\
&&+\, \frac{f_\pi^2}{4} {\rm Tr}\left(\nabla_\mu \Sigma^\dagger
\nabla^\mu \Sigma\right)
+ \frac{f_\pi^2}{4} m_\pi^2 {\rm Tr}(\Sigma + \Sigma^\dagger)
\label{chiralag2} \eea
and the corresponding nucleon propagator $S_F(\mN v+k)$ is
\be
S_F(\mN v+k) = \frac{1}{v\cdot k + i 0^+}\, .
\ee
As mentioned, the HFF is based on simultaneous expansion
in the chiral parameter and in ``$1/\mN$".
We have so far considered only the leading-order terms in $1/\mN$, {\it i.e.},
$O((1/\mN)^0)$. The
$\frac{1}{\mN}$ corrections can be gotten directly
from the Lagrangian (\ref{chiralag})\cite{grin,pmr1}:
\bea
{\cal L}_{\frac{1}{\mN}} &=&
\frac{1}{2\mN}  \Bbar \left( - D^2 + (v\cdot D)^2 + S^{\mu\nu} \Gamma_{\mu\nu}
+ \gA^2 (v\cdot \Delta)^2
+ 2\gA \left\{v\cdot \Delta, S\cdot D\right\}\right) B
\label{oneoverm}
\eea
where
$\Gamma_{\mu\nu}= \frac{\tau_a}{2} \Gamma_{\mu\nu}^a
=\del_\mu \Gamma_\nu -\del_\nu \Gamma_\mu + [\Gamma_\mu, \Gamma_\nu]$.

While eq.(\ref{oneoverm}) is the first ``$1/\mN$" correction, it is not the
entire
story to the order considered. One can see that it is also the next order in
the chiral counting in derivatives and  expected in any case
independently of the inverse baryon mass expansion.
Generally, the coefficient appearing in each term should be viewed as a
parameter rather than as a constant to be fixed by chiral symmetry in
HFF.
The most general chiral Lagrangian of order $\O(Q)$ relative to the
leading-order terms is
\be
{\cal L}_1 =
\frac{1}{2\mN}  \Bbar \,\Gamma_{1/m}\, B
- \frac{b_9}{2 \mN} \left(\Bbar B\right)\,
        \left(\Bbar i S\cdot\Delta B\right)
- \frac{b_{10}}{2 \mN} \left(\Bbar S^\mu B\right)\,
        \left(\Bbar i \Delta_\mu B\right)
\label{Lag1}
\ee
with
\bea
\Gamma_{1/m} &=&
- D^2 + (v\cdot D)^2 + (1+b_1) S^{\mu\nu} \Gamma_{\mu\nu}
+ (\gA^2 +b_2) (v\cdot \Delta)^2
\nonumber \\
&& \ \ \ \
+\ 2(\gA + b_3) \left\{v\cdot \Delta, S\cdot D\right\}
+ b_4 \Delta\cdot\Delta
+ 2 b_5 \left[v\cdot \Delta, S\cdot D\right]
\nonumber \\
&& \ \ \ \
+\ b_6 m_\pi^2 \Tr(\Sigma + \Sigma^\dagger)
+ b_7 m_\pi^2 (\Sigma + \Sigma^\dagger)
- \frac{i}{2} b_8 S^{\mu\nu}
  (\del_\mu {\cal V}_\nu^S - \del_\nu {\cal V}_\mu^S),
\label{Gamma1m}\eea
where ${\cal V}_\mu^S$ denotes the external isoscalar vector field
and $b_i$ $(i=1,2,\cdots,10)$ denotes dimensionless constants.
We have written down this part of Lagrangian for completeness although
it turns out that none of the terms in (\ref{Lag1}) contribute
to the process we are interested in to the chiral order considered.
We have checked that there is no $1/\mN$ correction in our calculation.
One might think that there would be non-zero contributions to
$V^i$ or $A^0$ coming from the contact terms that
have $S\cdot D$ or $v\cdot \Delta$, but these terms are forbidden
by parity.

\subsection{Renormalization}
\label{renormalization:S}
\indent

As in \cite{pmr1}, we shall adopt the dimensional regularization scheme
to handle ultraviolet singularities in loop calculations.
It has the advantage of avoiding
power divergences like $\delta (0)\sim \Lambda^4_{cut}$ where
$\Lambda_{cut}$ is the cut-off mass.
In $d= 4 - 2 \epsilon$ dimensions,
all the infinities appearing in one-loop calculations
are absorbed in $\frac{1}{\epsilon}$.
To define our conventions and
to apply the procedure to the case of the exchange vector currents,
we repeat some of the essentials of ref.\cite{pmr1}.

Let us begin with the bare Lagrangian
\be
\zero{\L} = \zero{\L}_0 + \zero{\L}_1 + \zero{\L}_2 +
\cdots\label{barelag1}
\ee
Here the subscript denotes the chiral power ${\bar \nu}_i$
as given by eq.(\ref{counting}). Note that
the zeroth-order term $\zero{\L}_0$ is the same with the Lagrangian
(\ref{chiralag2}) with the replacement,
$B\rightarrow \zero{B}= \sqrt{Z_N} B$,
$\pi^a\rightarrow \zero{\pi^a}= \sqrt{Z_\pi} \pi^a$
and all the couplings and masses replaced by their bare quantities.
The wavefunction renormalization constants are
\bea
Z_N &=& 1 + \frac{3 (d-1) \gA^2 \qdiv}{4 f_\pi^2},
\\
Z_\pi &=& 1 - \frac{2 \qdiv}{3 f_\pi^2}.
\eea
The renormalized quantities of interest are related to the bare ones by
\bea
\zero{m}_\pi^2 &=& m_\pi^2
\left[1 + \frac{\qdiv}{2 f_\pi^2}\right],
\nonumber \\
\zero{f}_\pi &=& f_\pi
\left[1 - \frac{\qdiv}{f_\pi^2}\right],
\nonumber \\
\zero{g}_A &=& g_A
\left[1 - \left(1 + \frac{d g_A^2}{2}\right)
\frac{\qdiv}{f_\pi^2}\right]
\eea
where $\qdiv$ and $\ldiv$ are, respectively, the quadratic
and logarithmic divergences,
\bea
\Delta(m_\pi^2)&=&
- \frac{m_\pi^2}{16\pi^2} \Gamma(-1+\epsilon)
\left(\frac{m_\pi^2}{4\pi\mu^2}\right)^{-\epsilon},
\label{DM2}\\
\ldiv &=& \frac{1}{16\pi^2}
\Gamma(\epsilon) \left(\frac{m_\pi^2}{4\pi\mu^2}\right)^{-\epsilon}
\label{LL}
\eea
with $\mu$ the renormalization scale.

For nuclear applications, we need to add four-Fermi interactions to
(\ref{barelag1}). The relevant bare Lagrangian can be written as
\bea
\zero{\L}_{\rm 4pt} =
- \frac{\mu^{4-d}}{2} \left[
  \zero{C}_S^S \left(\zero{\Bbar} \zero{B}\right)^2
  + \zero{C}_S^V \left(\zero{\Bbar} {\vec \tau} \zero{B}\right)^2
  -\ 4\, \zero{C}_T^S \left(\zero{\Bbar}S_\mu \zero{B}\right)^2
  - 4\, \zero{C}_T^V
  \left(\zero{\Bbar}S_\mu {\vec \tau} \zero{B}\right)^2
\right]\label{fourfermi}
\eea
with
\bea
\zero{C}_S^S &=& -\frac{3(d^2-1) \gA^4}{32 \fpi^4} \qdiv + C_S^{S},
  \nonumber \\
\zero{C}_S^V &=& - \left[-\frac18 +\frac{1-d}{4}\gA^2
  + \frac{d^2-1}{16} \gA^4\right]\frac{\qdiv}{\fpi^4}
  + C_S^{V},
  \nonumber\\
\zero{C}_T^S &=& \left[1-3(d-2) \gA^2\frac{\qdiv}{\fpi^2}\right] C_T^{S},
  \nonumber \\
\zero{C}_T^V &=& \left[1-3(d-2) \gA^2\frac{\qdiv}{\fpi^2}\right] C_T^{V}.
\eea
We can easily see that the ``counter terms" $C_T^S$ and $C_T^V$ are
not necessary for removing divergences from one-loop corrections to
the four-Fermi Lagrangian: There are no divergences involving $S_\mu$.
Such terms however could arise when one integrates out heavy-meson
degrees of freedom.
In the heavy-meson-exchange picture, the coefficients
$C_S^S$ and $C_T^S$ would come from the exchange of
isoscalar mesons while $C_S^V$ and $C_T^V$ from that of
isovector mesons. The coefficients $C_S^S$ and $C_S^V$ would come from
mesons of even intrinsic parity (such as scalars and
vectors) while $C_T^S$ and $C_T^V$ would result from odd intrinsic
parity mesons (such as pseudoscalars and axial-vectors).
In coordinate space, the Lagrangian (\ref{fourfermi}) gives zero-ranged
interactions, so in two-nucleon systems, only two terms effectively
contribute.  Weinberg\cite{wein90} chose $C_S^S$ and $C_T^S$
($C_S$ and $C_T$ in his notation) while in our calculations,
we take $C_S^S$ and $C_S^V$.\footnote{Due to the
antisymmetry of the two-body wavefunction and since the force is zero-ranged,
the contact Lagrangian can be rewritten as
\be
V_{\rm 4pt}(\bm r)=
\delta(\bm r)\left[(C_S^S-C_T^S-C_S^V-3 C_T^V) + 2 P_\tau (C_S^V-C_T^S)
\right],\ \ \ \ P_\tau=\frac{\tau_1\cdot\tau_2 + 1}{2}.
\ee
Thus, we can convert the Weinberg parameters $(C_S,\ C_T)$ into our
parameter $(C_S^S,\ C_S^V)$ by $C_S^S= C_S - 2 C_T$ and $C_S^V= -C_T$.}

There are many counter-terms that can enter in  ${\cal L}_2$ at
order $\O(Q^2)$ compared to the leading chiral order terms. Some of them
were written down in \cite{pmr1} for the exchange axial currents. Here, we
list only those terms that are needed for exchange vector currents:
\bea
{\cal L}_2 &=&
  i \gA \frac{c_2}{\fpi^2} \Bbar \, S^\mu v^\nu v^\alpha
    \left(\Delta_\mu D_\nu D_\alpha
  - \stackrel{\leftarrow}{D_\nu} \Delta_\mu D_\alpha
  + \stackrel{\leftarrow}{D_\nu} \stackrel{\leftarrow}{D_\alpha}
    \Delta_\mu \right) B
\nonumber \\
&+&
  \frac{\gA^3}{2 \fpi^2} c_7 \Bbar \epsilon^{\mu\nu\alpha\beta} v_\mu \left\{
  \Delta_\nu, \Gamma_{\alpha\beta}\right\}B
\nonumber \\
&+&i\frac{\gA}{\fpi^2} c_8 \Bbar
  v^\nu v^\alpha S^\beta \left[\Delta_\nu, \Gamma_{\alpha\beta}\right] B
\nonumber \\
&+& \left(\frac{i\,\gA c_9}{4 \fpi^2}\,
 \Bbar \,\Delta^a_\mu {\cal P}^{\mu\nu} \Gamma_{\nu\alpha}^b S^\alpha
 \Lambda_{ab} \,B + \mbox{h.c.}\right)
\nonumber \\
  &-& {G}_1\, \Bbar \Gamma_{\mu\nu} S^{\mu\nu} B \, \Bbar B
  -{G}_2\, \Bbar \Gamma_{\mu\nu} B \, \Bbar S^{\mu\nu} B
\nonumber \\
&-& \frac{1}{2}\, E_T^V\,
  D_\alpha(\Bbar\tau^a S^{\mu\alpha} \, B)
   \, D^\beta(\Bbar\tau^a S_{\mu\beta} \, B)
\label{counterterm}\eea
where $D_\mu (\Bbar X B) \equiv \del_\mu (\Bbar X B) + \Bbar\,
\left[X, \,\Gamma_\mu\right] B$ for any $X$,
$\Gamma_{\mu\nu}= \frac{\tau_a}{2} \Gamma_{\mu\nu}^a
=\del_\mu \Gamma_\nu -\del_\nu \Gamma_\mu + [\Gamma_\mu, \Gamma_\nu]$ and
\bea
\Lambda_{ab} &=& \delta_{ab} - \frac13 \tau_a \tau_b
= \frac23 \delta_{ab} - \frac{i}{3} \epsilon_{abc}\tau_c,
\\
\calP^{\alpha\beta} &=& v^\alpha v^\beta - g^{\alpha\beta}
- \frac43 S^\alpha S^\beta.
\eea
All the divergences encountered in our calculation can be removed
by
\bea
c_2 &=& \frac{d-3}{3} \gA^2 \ldiv + c_2^R,
\nonumber \\
c_7 &=& -\ldiv + c_7^R,
\nonumber \\
c_8 &=& 2 \ldiv + c_8^R,
\nonumber \\
\frac{G_1 + G_2}{2} &=&
- \frac{\gA^2}{4 \fpi^4} \ldiv\,
\left[1 - (d-1) \gA^2 + 4 \fpi^2
  \left(C_S^S-C_S^V + C_T^S - C_T^V\right)\right] + G_S^R,
\nonumber \\
\frac{G_1 - G_2}{2}&=&
\frac{\gA^2}{4 \fpi^4} \ldiv\,
\left[1 - 4 \fpi^2 \left(C_S^S+C_S^V - C_T^S - C_T^V\right)\right]
+ G_A^R,
\nonumber \\
E_T^V &=& \frac{\gA^4}{2 \fpi^4}\,\ldiv + E_T^{V,R}.
\eea
Note the $c_2$ term has been introduced in \cite{pmr1} while the $c_7$
and $c_8$ terms are new. They are consistent with
Ecker's counter terms obtained by heat-kernel method \cite{Ecker:heat}.
The counter term $c_9$ is not associated with any divergence
but is needed if the $\Delta$-resonance degree of freedom is integrated
out as we will do in our calculation.
We also note that
$c_i^R$, $G_i^R$, $E_T^{V,R}$ and $c_9$ are all finite,
$\mu$-independent renormalized constants.

\subsection{Chiral perturbation theory in nuclear processes}
\label{nuclear:S}
\indent

We wish to calculate operators effective in nuclei for transitions
induced by the vector and axial-vector currents
denoted respectively by $V_\mu$ and $A_\mu$ associated with the electroweak
fields ${\cal V}_\mu$ and ${\cal A}_\mu$.
As mentioned, there are subtleties in applying ChPT
to nuclear processes. Firstly, we need realistic
initial and/or final nuclear wavefunctions. This is not a trivial matter
in quantum field theory in general and in ChPT in particular.
Perturbation series to bound states fail to converge because
they consist of infinite number of ``reducible"
ladder diagrams which are of same chiral order.
As stressed by Weinberg, chiral perturbation theory is useful in
nuclear physics {\it only for ``irreducible" diagrams} that are
free of infrared divergences.
This means that both in nuclear forces and in exchange currents,
reducible graphs are to be taken care of by a Schr\"{o}dinger equation
or its relativistic generalization with irreducible graphs entering
as potentials.
This also implies that in calculating exchange currents in ChPT,
we are to use the wave functions so generated to calculate matrix elements
to obtain physical amplitudes. This is of course the standard practice in
the theory of meson-exchange currents but it is also in this sense that
ChPT is predictive in nuclei. We should stress that
this precludes what one might call
``fully consistent chiral perturbation theory" where nuclear forces,
nuclear currents and wave functions are all calculated to the same order
of chiral perturbation expansion.  Such a calculation even if feasible
is likely to make no sense unless one learns to solve QCD for nuclei.

Let $M$ be the transition matrix of the
operator $\M$
\be
M= \langle \Psi_f | \M | \Psi_i \rangle
\ee
where $|\Psi_i\rangle$ and $|\Psi_f\rangle$ are the
initial and final wavefunctions, respectively.
The $\M$ in ChPT is consistently expanded as
\be
\M=\M_{-2} +\M_{-1} +\M_{+1} + \M_>\label{mu}
\ee
where the subscript on $\M$ is $\nu$ for $Q^\nu$ and
$\M_>$ contains $O(Q^2)$ and higher in chiral expansion.
Our claim that will be substantiated below is
that $\M_>$ is small and hence can be ignored.
This is guaranteed by the higher power in $Q$ and
by the ``naturality" argument.
The power of ChPT is that we can actually calculate all first three
terms {\em accurately} if we have accurate wave functions and {\em accurate}
experimental information from other experiments, these
to fix the counter terms appearing in the vertices.
Since we ignore higher order terms, the terms retained
fail to satisfy the exact continuity equation implied by
the potential that gives rise to the {\it accurate} wave functions.
But those ``short-range" terms (and multi-body terms that figure
in many-body systems) that must enter into
$\M_>$ to satisfy the continuity equation is suppressed {\it numerically} by
some order in the chiral counting and if the correction is small, then we
are free to
ignore them even though by so doing we will be violating certain Ward
identities.\footnote{This means that we are satisfying Ward identities
-- associated
with conserved currents -- to the chiral order we retain and the
violation will occur only at the next chiral order that is ignored.
We note that this is a
point frequently missed in the literature.}
If the chiral expansion reproduces QCD at some low energy and
if the wave function is accurate enough, then the low-order expansion
in the current must be arbitrarily accurate at that low energy.
What we have accomplished in this paper
is that with the wave function chosen,
the chiral expansion converges for the thermal capture process,
leaving errors of order 0.6~\%! This means that the short-ranged term
required by the continuity equation lies within that error bar.

Let us consider many-body systems as in the case of
axial-charge transitions \cite{pmr1}.
To the next-to-next-to-leading order as in
(\ref{mu}), chiral symmetry again says that at
$E\ll \Lambda_\chi$, three-body and
four-body etc currents and forces are zero. If one were
to build an arbitrary, though phenomenologically guided, model unconstrained
by chiral symmetry, there would be
no reason why one should not include three-body, four-body
etc. currents and forces and/or what might correspond to
higher chiral order terms
although he/she might find them small
{\it a posteriori}. But then to satisfy one's continuity equation, one would
have to keep adding a chain of what one might call ``higher-order morass."
What chiral symmetry says is that
this is not an efficient way of doing physics
if it has anything to do with QCD.

In calculating loop graphs involving two-pion exchange,
we encounter divergences that can be removed by four-Fermi counter terms
and their second derivatives.
When Fourier transformed into coordinate space,
these terms appear as zero-ranged functions --
$\delta^3({\bm r})$ -- and their derivatives, all
with renormalized coefficients.
The prescription needed for handling
these short-ranged terms
does not follow from chiral symmetry alone
and will have to be justified on a more general ground.
As was explained in detail in \cite{pmr1},
consistency with ChPT demands that those terms in exchange
currents be {\em (a) negligible in magnitude  and (b) additionally
suppressed by nuclear correlations when embedded in nuclear medium.}
To give an example, low-mass vector mesons
$(\omega, \rho)$ cannot contribute to the
time part of the exchange axial current and the
exchange magnetic moment operator while they do to nuclear forces.

\section{Exchange Currents}
\indent

Let us focus on the vector currents $V^\mu(k;q)$ involving two nucleons
with the kinematics defined by
$$
N_1(p_1) + N_2(p_2) \rightarrow N_1(p_1') + N_2(p_2') + {\cal V}^\mu(k),
$$
where $k^\mu$ is the momentum carried by outgoing
external vector field ${\cal V}^\mu$ and
$q^\mu$ is the transferred four-momentum between nucleons
$q^\mu = \frac12 (p_2' - p_2 + p_1 - p_1')$.
We are interested in the space component of the vector current
with $k^\mu\rightarrow 0$. The process (\ref{np})
that we will analyze in detail
involves the magnetic moment operator (MMO)
\be
{\bm \mu}({\bm r}) = e\,
\int\!\!\frac{d^3{\bm q}}{(2\pi)^3}\, \e^{i{\bm q}\cdot{\bm r}}
\,\,\left[ \frac{i}{2} {\bm \nabla}_{\bm k}
\times {\bm J}({\bm k}; {\bm q}) \right]_{{\bm k}={\bm 0}},
\label{muRmur}\ee
where $J^\mu = V^{3,\mu} + \frac12 B^\mu$, $B^\mu$ is the baryonic
(iso-scalar Lorentz-vector) currents.
Hereafter, we will denote $V^{3,\mu}$ simply by $V^\mu$.
We see from this formula that only terms linear
in $k^\mu$ can give nonvanishing contributions to the MMO.
\subsection{One-body (impulse) contribution}
\indent

In the long wavelength limit, the one-body vector current is dominated by the
MMO which is entirely determined by experiments:
\be
{\bm \mu}_{\rm{1B}} =
\frac{e}{2 m_p} ( \mu_p {\bm \sigma}_p + \mu_n {\bm \sigma}_n)
= \frac{e}{4 m_p} \left(
\mu_S \, \sum_i \,{\bm \sigma}_i
+ \mu_V\, \sum_i \,\tau^z_i \, {\bm \sigma}_i \right)
\label{mu1B}\ee
where $\mu_S \equiv \mu_p + \mu_n \simeq 0.87981$
and $\mu_V\equiv \mu_p-\mu_n \simeq 4.70589$.
As indicated in Table 1,
the corresponding space part of the one-body vector currents
is of order ${\cal O} (Q^{-2})$
in the chiral counting.
The meaning of this operator in the context of ChPT is as follows.
To leading order in the chiral expansion, the dominant contribution is
the sum of single-particle transitions with vertices modified by chiral
loop corrections. The renormalized constants (calculated to the chiral
order we shall consider) can then be fixed by experiments. This means that
there is no prediction involved at the level of the impulse approximation.
Given the single-particle operator, prediction is made in two-body and
higher-body corrections that can be computed in ChPT.
\subsection{Two-body contributions}
\indent

\begin{figure}
\centerline{\epsfig{file=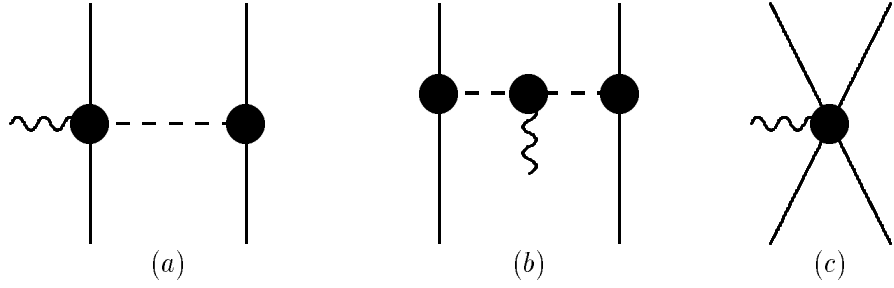}}
\caption{Generic graphs contributing to the exchange vector current.
The large filled circles
represent one-nucleon and one-pion irreducible graphs
and the solid (dashed) lines
renormalized nucleon (pion) propagators.
The graph $(a)$ and $(b)$ are the usual ``seagull"
and ``pion pole" graphs with one-loop correction, respectively,
and the graph $(c)$ represents one-pion and one-nucleon
irreducible graphs up to one-loop order.
}
\label{GeneV}
\end{figure}

The Feynman graphs for two-body exchange vector currents are
given in Fig.~\ref{GeneV}. We shall now discuss their contributions
in the ascending chiral order.

\subsubsection{Tree contributions}
\indent

\begin{figure}[b]
\centerline{\epsfig{file=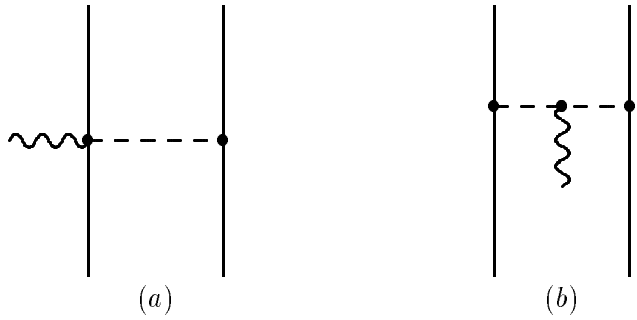}}
\caption{Tree graphs contributing to the exchange vector current,
the ``seagull" $(a)$ and the ``pion pole" $(b)$ graph.}
\label{treeV}
\end{figure}

It follows from the counting rule eq.(\ref{counting}) that
the leading ($O(Q^{-1}$) or $O(Q)$ relative to the impulse term)
two-body currents come from the tree graphs, Fig.~\ref{treeV},
with the vertices given by the bare ones (which will be renormalized
at higher order and identified with the experimental values).
Up to leading chiral order, the four-Fermi contact terms cannot
contribute to the vector currents \cite{mr91}.
The leading two-body currents therefore are
\bea
V^\mu_{tree}(a) &=& i\ttz \frac{\gA^2}{\fpi^2} S_1^\mu
\frac{S_2\cdot q_2}{\mpi^2-q_2^2} + (1\leftrightarrow 2),
\nonumber\\
V^\mu_{tree}(b) &=& -2\,i \ttz \frac{\gA^2}{\fpi^2}
\frac{S_1\cdot q_1}{\mpi^2-q_1^2}\,
\frac{S_2\cdot q_2}{\mpi^2-q_2^2}\, q^\mu
\label{Vtree}
\eea
where $q_i= p_i'- p_i$ is the momentum transferred to the $i$th nucleon,
$q_1^\mu = -q^\mu -\frac12 k^\mu$ and
$q_2^\mu = q^\mu -\frac12 k^\mu$.
The coupling constants and masses in the above equation
are renormalized (measured) constants
$g_A=1.257$ and $f_\pi=93$ MeV. The resulting MMO operator
denoted ${\bm \mu}_{tree}$, is (from eqs.(\ref{Vtree},\ref{muRmur}))
\be
{\bm \mu}_{tree} = e\,\frac{\gA^2 \mpi}{8\fpi^2}
\left\{\TtimesS \left[\frac23 y_1(x_\pi)- y_0(x_\pi) \right]
- \TtimesT \, y_1(x_\pi)\right\},
\label{murtree}
\ee
where
\bea
y_0(x) &=& \frac{\e^{-x}}{4\pi x},
  \nonumber \\
y_1(x) &=& \frac{\e^{-x}}{4\pi x} (1+ x) \ = \ - x\frac{d}{dx} y_0(x),
  \nonumber \\
y_2(x) &=& \frac{\e^{-x}}{4\pi x} \left(1+ \frac{3}{x} + \frac{3}{x^2}\right)
  \ = \ x\frac{d}{dx} \frac{1}{x} \frac{d}{dx} y_0(x)
\label{y012}\eea
and
\bea
{\hat T}_S^{(\odot)} &=& (\tau_1 \odot \tau_2)^z
({\bm \sigma}_1 \odot {\bm \sigma}_2),
 \nonumber \\
{\hat T}_T^{(\odot)} &=& (\tau_1 \odot \tau_2)^z \left[{\hat r}\,{\hat r}\cdot
({\bm \sigma}_1 \odot {\bm \sigma}_2)
-\frac13 ({\bm \sigma}_1 \odot {\bm \sigma}_2)\right]
\label{TST}\eea
with $\odot= \pm,\, \times$.
We should mention that this is the same as what one gets from
the corresponding ``pair" and ``pionic" currents obtained in \cite{chemrho}.
The only thing new here is its precise place in the chiral expansion.

\subsubsection{Loop corrections to the one-pion-exchange currents}
\indent

\begin{figure}
\centerline{\epsfig{file=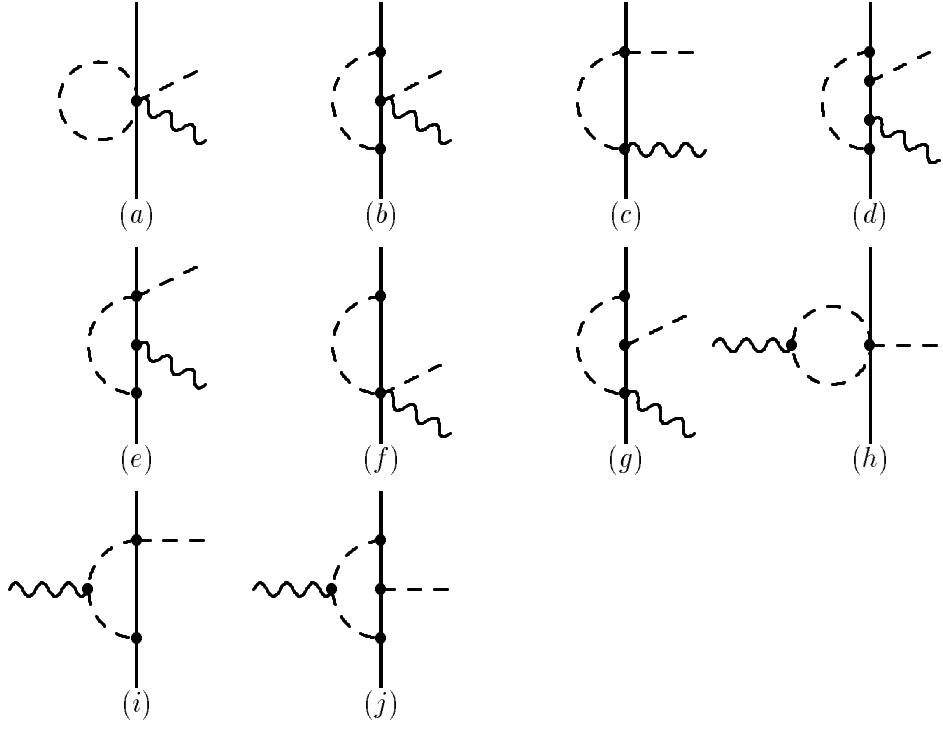}}
\caption{
One-nucleon, one-pion irreducible one-loop graphs contributing
to the four-point ${\cal V} \pi NN$ vertex.
}
\label{feyFour}
\end{figure}

\begin{figure}
\centerline{\epsfig{file=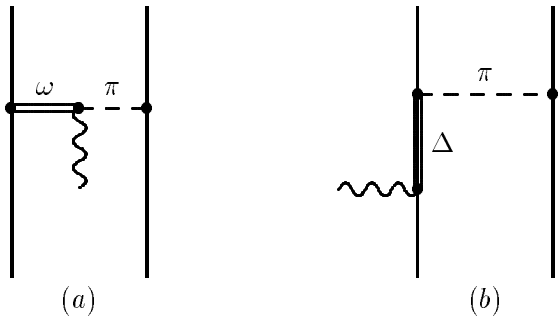}}
\caption{
Graphs that correspond to the full one-loop corrections to the
one-pion-exchange graphs.
These graphs plus the tree graphs are referred to as
``generalized tree" graphs.}
\label{GTO}
\end{figure}

The first one-loop corrections corresponding to $O(Q^2)$
relative to the leading-order tree discussed above are renormalizations
of the vertices in one-pion exchange graphs.
We shall first show that the corrections appear
only at the ${\cal V} \pi NN$ vertex.
The $\pi NN$ proper function up to one-loop
accuracy is \cite{pmr1}
\be
\Gamma_{\pi NN,i}^a(q) = - i \tau_i^a \frac{g_A}{f_\pi}\,
 q\cdot S_i
\ee
where terms proportional to $v\cdot q$ are higher-order
and hence can be neglected. There are thus
no one-loop corrections to the $\pi NN$ vertex.
Now the Ward identity restricts
the ${\cal V} \pi \pi$ proper function to the form
\bea
\Gamma_{{\cal V} \pi\pi}^{\mu,ab}(q_1,q_2)
&=& 2i\epsilon_{3ab} \left\{
q^\mu + \left(q^\mu-\frac{k^\mu\, k\cdot q}{k^2}\right) \left[
F_{\pi\pi}^V(k^2)-1\right]\right\}
\nonumber \\
&=& 2i\epsilon_{3ab}\, q^\mu + \O(k^2)
\eea
where $F_{\pi\pi V}^V(k^2)$ is the electromagnetic form factor of the pion,
$F_{\pi\pi}^V(k^2) = 1 + {\cal O}(k^2)$.
This shows that there are no one-loop corrections from this vertex
to the magnetic moment operator.
Therefore we are left with
the ${\cal V} \pi NN$ vertex appearing in
Fig.~\ref{GeneV}$a$.
Here the situation is quite analogous to
the one-loop corrections in the exchange axial-charge currents
\cite{pmr1}.
The one-loop Feynman graphs contributing to the
relevant vertex are given in Fig.~\ref{feyFour}.
These one-loop graphs are computed in
Appendix~\ref{VpiNN:A}. The result can be summarized simply as follows.
The key point to underline is that for the magnetic moment operator,
there are no corrections from the
finite loop terms. The only contribution comes from the finite counter terms
that describe the degrees of freedom that are integrated out from
our low-energy effective Lagrangian. In fact there are two such terms:
\bea
{\cal L}_{\mbox{\tiny CT}} &=&
  \frac{\gA^3\,c_7^R}{4 \fpi^2}\,
  \Bbar \epsilon^{\mu\nu\alpha\beta} v_\mu \,
  \Delta_\nu^a \Gamma_{\alpha\beta}^a\,B
\nonumber \\
&+& \left[\frac{i\,\gA\, c_9}{4 \fpi^2}\,
 \Bbar \,\Delta^a_\mu
  {\cal P}^{\mu\nu} \Lambda_{ab}
  \Gamma_{\nu\alpha}^b S^\alpha \,B + \mbox{h.c.}\right]\label{counter}
\eea
where
$\Gamma_{\mu\nu}= \frac{\tau_a}{2} \Gamma_{\mu\nu}^a$
and $\Delta_\mu=\frac{\tau_a}{2} \Delta_\mu^a$.
Here we have defined the spin-$\frac32$ and isospin-$\frac32$ projection
operators by
$${\cal P}^{\mu\nu}=
  \left(v^\mu v^\nu - g^{\mu\nu} - \frac43 S^\mu S^\nu\right),
\ \ \ \
\Lambda_{ab}=
  \left(\delta_{ab} - \frac{\tau_a \tau_b}{3}\right).
$$
The resulting one-pion exchange current, denoted
$V_{1\pi}^\mu$, is of the form
\bea
V_{1\pi}^\mu &=& -i \tau_2^z\, \frac{\gA^4}{2 \fpi^4}\,c_7^R\,
\frac{q_2\cdot S_2}{\mpi^2-q_2^2}
\, \epsilon^{\mu\nu\alpha\beta}
v_\nu k_\alpha q_\beta
\nonumber\\
&-&
\frac{\gA^2}{4 \fpi^4}\,c_9\,
\frac{q_2\cdot S_2}{\mpi^2-q_2^2}\,
\left[
\left(\frac23 \tau_2^z - \frac{i}{3} \ttz\right)
\,q_{2 \beta} \calP_1^{\beta\alpha}
(k_\alpha\, S_1^\mu - g_\alpha^\mu\,k\cdot S_1)
\right.\nonumber \\
&&\ \ \ \ \ \ \ -\ \left.
\left(\frac23 \tau_2^z + \frac{i}{3} \ttz\right)
\,(S_1^\mu\,k_\alpha - g^\mu_\alpha\,k\cdot S_1)
\calP_1^{\alpha\beta} q_{2 \beta} \right]
\nonumber \\
&+& (1\leftrightarrow 2).
\label{VmuVector}\eea
The counter terms bring in two parameters, $c_7^R$ and $c_9$, to be fixed.
These should be determined from experiments.
Relevant experiments would be isovector pion photoproduction
or radiative weak amplitudes at low energy.
However, the presently available data are not accurate enough
to fix the constants sufficiently reliably.
The difficulty lies in the fact that
those processes are dominated by the seagull term and nucleon pole term
and the constants we need to fix represent small corrections that cannot
be extracted from the data. We shall therefore estimate them
by the resonance saturation in a way analogous to what is done for the
$O(Q^4)$ Lagrangian counter terms for $\pi\pi$ scattering \cite{ecker}.
The relevant resonances are
the $\omega$ and $\Delta$ saturating, respectively,
the constants $c_7^R$ and $c_9$.
The results are
\bea
{\bar c}_\omega &\equiv& \frac{\gA^2 \mpi^2}{\fpi^2}\, c_7^R
= \frac{g_\omega^2 \mpi^2}{8 \pi^2 \gA (m_\omega^2 - \mpi^2)}
\simeq 0.1021,
\nonumber \\
{\bar c}_\Delta &\equiv& \frac{2 \mpi^2}{9 \fpi^2}\, c_9
= \frac{2\, \mu_T\, {\cal C}\, \mpi^2}{9 \gA (m_\Delta-m_N) m_N} \simeq 0.1667
\eea
where $g_\omega$ is determined from the $\omega\rightarrow \pi\gamma$ decay,
$g_\omega=17.55$, and the $N\Delta$ transition magnetic moment $\mu_T$ and
the $\pi N \Delta$ coupling ${\cal C}$ come from the fit to the
$\Delta$ properties as explained in \cite{jenkinsetal}, ${\cal C}=-1.73$ and
$\mu_T=-7.7\pm 0.5$. For details, see Appendix~\ref{Res:A}.
The resulting magnetic moment operator that we will use later is
rather simple:
\be
{\bm \mu}_{1\pi}
= {\bm \mu}_{1\pi}^\omega + {\bm \mu}_{1\pi}^\Delta
\ee
with
\bea
{\bm \mu}_{1\pi}^\omega &=& -e\,\frac{\gA^2 \mpi}{8 \fpi^2}
  {\bar c}_\omega \left\{\left(\TplusS+\TminusS\right)
  \frac{{\bar y}_0(x_\pi)}{3}
   + \left(\TplusT+\TminusT\right)\,y_2(x_\pi) \right\},
\nonumber \\
{\bm \mu}_{1\pi}^\Delta &=& -e\,\frac{\gA^2 \mpi}{8 \fpi^2} {\bar c}_{\Delta}
  \left\{\left(\TplusS+\TminusS-\TtimesS \right)
  \frac{{\bar y}_0(x_\pi)}{3}
   + \left(\TplusT+\TminusT+\frac12\TtimesT\right)\,y_2(x_\pi) \right\}.
\nonumber\\
\label{mur1pi}\eea
This operator is equivalent to Fig.~\ref{GTO} calculated in the conventional
meson-exchange current study. What is significant -- and novel --
in the context of chiral perturbation theory is that
chiral symmetry constrains that there be no other contribution to
the next-to-leading order one-pion exchange magnetic moment operator:
{\it There are no chiral loop corrections to the one-pion-exchange
operator}. This is
in contrast to the case of axial-charge transitions \cite{pmr1} where
chiral loop corrections to the one-pion exchange axial charge operator
play a relatively important role.
This suggests to
lump all four terms of Fig.~\ref{treeV} and Fig.~\ref{GTO}
together and call them ``generalized tree operators".

We should explain briefly
why there are no genuine loop corrections
(``chiral log terms") here in contrast to the
axial-charge operator of \cite{pmr1}
and how the ``counter terms" eq.(\ref{counter})
are saturated by the resonances.
The genuine loop corrections arise from $q^2$-dependent
terms where $q^\mu$ is the momentum transferred by the pion.
Now the nucleon propagators in a loop cannot
bring in a $q^2$-dependence because of the {\em softness}
of the momentum and $v\cdot q=0$. This means that if there is any
momentum-dependent term, it must
arise only from the momentum dependence of the pion
propagators in the loop.  We see from Fig.~\ref{feyFour} that
the first seven graphs $(a-g)$ are momentum-independent and
the next three graphs $(h,i,j)$ are $k^2$-dependent but $q^2$-independent.
Therefore if any,
chiral log terms can occur only at two-loop (or higher) order.

\subsubsection{Two-pion-exchange currents}
\indent

\begin{figure}
\centerline{\epsfig{file=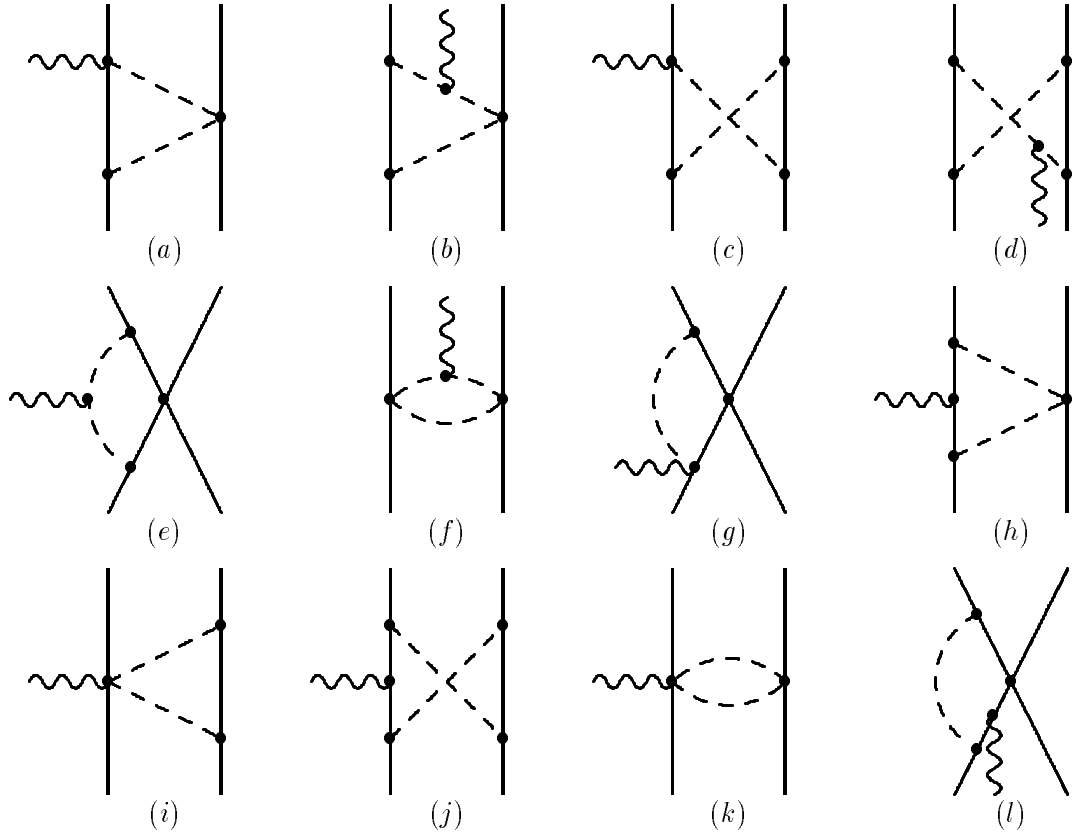}}
\caption{One-pion and one-nucleon irreducible one-loop graphs
contributing to the two-body vector current.
Note that one-loop graphs which contain purely nucleonic
intermediate state are excluded.
Only the first four graphs ($a$, $b$, $c$ and $d$) contribute
to the magnetic moment operator.}
\label{TwoV}
\end{figure}

There are numerous diagrams of genuine loop character that can
contribute in general kinematics to the two-pion-exchange two-body
current. However a drastic simplification is obtained for the magnetic
moment operator in heavy-fermion formalism:
only four graphs $(a)$, $(b)$, $(c)$ and $(d)$
of Fig.~\ref{TwoV} give non-vanishing contributions to the MMO.
The graph $(e)$ gives only zero-ranged operator and so will vanish in the
matrix element,
the graph $(f)$ is non-zero but cannot contribute to the MMO as
it does not contain terms linear in $k^\mu$,
the graph $(g)$ is identically zero because it is proportional
to $v\cdot S=0$ and the remaining graphs
($(h)$, $(i)$, $(j)$, $(k)$ and $(l)$),
being proportional to $v^\mu$, do not contribute
to the space component of the vector current.
The calculation of the graphs is described in
Appendix~\ref{Jem:A}.

We pause briefly to explain why the zero-ranged
graph $(e)$ does not contribute in our calculation.\footnote{
There are exceptional cases
where this argument breaks down. This matter is discussed in section 5.}
Such zero-ranged two-pion-exchange graphs could be divergent
(that is, they could have the $\frac{1}{d-4}$ pole).
This divergence can be removed by a
counter term generated by $\L_2$, leaving us a
zero-range term with a finite coefficient. As in the case
of the counter terms
$c_7^R$ and $c_9$, the (renormalized) coefficients
should be fixed by experiments. However there are no experimental data,
so we saturate them by resonances.
As shown explicitly in Appendix~\ref{Res:A},
the coefficients obtained by this procedure come out to be simply zero.
This is because the only low-lying resonance that can contribute
is the $\rho$-meson, but its exchange does not give rise to
terms linear in $k^\mu$, and so gives vanishing contribution to the MMO.
Even if the coefficients were non-zero, they would not contribute since
short-correlations would ``kill" them anyway {\it unless the exchanged
particle comes down below the chiral scale as in the case of the scalar
meson in medium discussed in section 5}.

The resulting two-pion exchange MMO takes the form
\bea
{\bm \mu}_{2\pi} &=&
- \frac{e\,\mpi^3\,\gA^2}{256\pi^2 \fpi^4}
\left[\frac23 \left(\TplusS-\TminusS\right)
-\left(\TplusT-\TminusT\right)\right]\, x_\pi\frac{d}{dx_\pi} k_0(x_\pi)
\nonumber \\
&+& \frac{e\,\mpi^3\,\gA^4}{256\pi^2 \fpi^4}\left\{
\frac23 \TplusS\left[ -2 k_0(x_\pi)+4 k_1(x_\pi)
+ x_\pi \frac{d}{d x_\pi} k_0(x_\pi) + 2 x_\pi \frac{d}{d x_\pi} k_1(x_\pi)
\right]
\right. \nonumber \\
&&\ \ \ \ \ \ \
+\ \frac23 \TtimesS \left[\frac12 k_0(x_\pi) - k_1(x_\pi)
+ x_\pi \frac{d}{d x_\pi} k_0(x_\pi)\right]
\nonumber\\
&&\ \ \ \ \ \ \
+\ \TplusT\left[4 k_{\rm T}(x_\pi) - x_\pi \frac{d}{d x_\pi} k_0(x_\pi)
- 2 x_\pi \frac{d}{d x_\pi} k_1(x_\pi)\right]
\nonumber \\
&&\left.\ \ \ \ \ \ \
+\ \TtimesT\left[2 k_{\rm T}(x_\pi) - x_\pi \frac{d}{d x_\pi} k_0(x_\pi)
\right]\right\}
\nonumber \\
&+& e\,\delta({\bm r})\left[ -\frac12 G_S^R\,\TplusS - \frac12 G_A^R\,\TminusS
-\frac18 E_T^{V,R}\,\TtimesS\right].
\eea
The functions $k_0(x_\pi)$, $k_1(x_\pi)$ and $k_{\rm T}(x_\pi)$ are
defined by
\bea
k_{0}(x_\pi) &\equiv&
\frac{1}{\mpi^3}
\int\frac{d^3{\bm q}}{(2\pi)^3}\, {\rm e}^{i \,{\bm q}\cdot{\bm r}}\,
\int_0^1 dz\, {\rm ln}\left[1+z(1-z) \frac{{\bm q}^2}{\mpi^2}\right],
\nonumber \\
k_{1}(x_\pi) &\equiv&
\frac{1}{\mpi^3}
\int\frac{d^3{\bm q}}{(2\pi)^3}\, {\rm e}^{i \,{\bm q}\cdot{\bm r}}\,
\int_0^1 dz \frac{z(1-z) {\bm q}^2}{\mpi^2 + z(1-z) {\bm q}^2},
\nonumber \\
k_{\rm T}(x_\pi) &\equiv&
\frac{1}{\mpi^3}
x_\pi \frac{d}{d x_\pi}\frac{1}{x_\pi}\frac{d}{d x_\pi}
\int\frac{d^3{\bm q}}{(2\pi)^3}\, {\rm e}^{i \,{\bm q}\cdot{\bm r}}\,
\int_0^1\!\! dz\, \frac{z^2\,\mpi^2}{\mpi^2 +z(1-z){\bm q}^2}.
\eea
Analytic solutions are available for
nonzero $x_\pi$ for the functions $k_{0,1}(x_\pi)$:
\bea
k_0(x) &=& -\frac{1}{\pi x^2} \bfK_1(2 x),
\nonumber \\
\frac{d}{dx} k_0(x) &=& \frac{1}{\pi x^2} \left(
  x \bfK_0(2 x) + 2 \bfK_1(2 x) + x \bfK_2(2 x)\right),
\nonumber \\
k_1(x) &=& -\frac{1}{\pi x} \bfK_0(2 x),
\nonumber \\
\frac{d}{dx} k_1(x) &=& \frac{1}{\pi x} \left(\bfK_0(2 x)
  + 2 x \bfK_1(2 x)\right)
\eea
where $\bfK_n(z)$ are the $n$th order
modified Bessel functions.

As we explained in several places, what we have to deal with is
the finite-ranged contribution.
For the $n+p\rightarrow d+\gamma$ process,
the contributions proportional to the $\TplusS$ or $\TplusT$
are zero. Dropping zero-ranged operators that vanish by short-range
correlations, the MMO
can be expressed in terms of the ``fundamental" constants $\gA$ and $f_\pi$,
\bea
{\bm \mu}_{2\pi} &=&
\frac{e\,\mpi^3\,\gA^2}{16\pi^2 \fpi^4}
\left(\frac23 {\hat T}^{(-)}_S -{\hat T}^{(-)}_T\right)\,
\int_1^\infty\!\!\!
  dt\, t \sqrt{t^2-1}\, y_1 (2 t x_\pi)
\nonumber \\
&+& \frac{e\,\mpi^3\,\gA^4}{16\pi^2 \fpi^4}
\int_1^\infty\!\!\! dt\,
\left\{
\frac23 {\hat T}^{(\times)}_S
\frac{t^3}{\sqrt{t^2-1}}
  \left(y_1(2 t x_\pi) -\frac12 y_0(2 t x_\pi)\right)
\right. \nonumber \\
&&\left.\ \ \ \ \ \ \
-{\hat T}^{(\times)}_T
\left( t \sqrt{t^2-1} \,y_1 (2 t x_\pi)
  - \frac{2 t^3 - t}{\sqrt{t^2-1}}\,y_2(2 t x_\pi)
     \right)
\right\}
\label{mur2pi}\eea
where $x_\pi= m_\pi r$ and the functions $y$'s are defined in
eq.(\ref{y012}).

In summary,
the total magnetic moment operator up to one-loop accuracy then is
\be
{\bm \mu}={\bm \mu}_{\rm 1B}
+{\bm \mu}_{tree} +{\bm \mu}_{1\pi}+{\bm \mu}_{2\pi}.
\label{mutotal}
\ee

\section{Application to $n+p\rightarrow d + \gamma$}
\indent

\subsection{A long-standing discrepancy}
\indent

The radiative $np$ capture $n+p \rightarrow d + \gamma$
at threshold is the most unambiguous nuclear process where the effect of
pionic currents can be clearly seen. We shall apply the magnetic
moment operator derived in chiral perturbation expansion to
this process.

The experimental cross-section measured accurately with the neutrons of
thermal velocity, $v_n = 2200\ \mbox{m/sec}$,
is \cite{cox}
\be
\sigma_{\rm{exp}} = (334.2 \pm 0.5)\ \mbox{mb}.\label{npexp}
\ee
We first describe the prediction made in impulse approximation.
For this, we note that the process involves very low energy,
the relative momentum in the center of mass system being
\be
p= \frac12 \mN v_n  \simeq 3.4451\times 10^{-3}\ \mbox{MeV}
\simeq \left(5.7278\times 10^{4}\ \mbox{fm}\right)^{-1}
\label{p_def}\ee
where the nucleon mass $\mN$ is defined by twice the reduced mass
of the proton and neutron,
$\mN = 2 m_p m_n/(m_p + m_n) \simeq 938.92$~MeV.
At this extremely low energy,
the process is predominantly governed by the isovector M1 transition operator.
We write the wavefunctions of the initial ${}^1 S_0$ $np$ state and
the final deuteron state, respectively, by
\bea
\Psi_{np} ({\bm r}) &=&
\frac{1}{\sqrt{4 \pi}\, r} \,u_0(r)\, \chi_{00}\, \zeta_{10},
\nonumber \\
\Psi_d ({\bm r}; J_z)&=& \frac{1}{\sqrt{4 \pi}\, r}
\left[u(r) + \frac{w(r)}{\sqrt{8}} \,S_{12}({\hat r})\right]
\,\chi_{1 J_z}\, \zeta_{00}
\eea
where $\chi_{J J_z}$ ($\zeta_{T T_z}$) is
the spin (isospin) spinor
and
\be
S_{12}(\hat{r}) \equiv
3\,{\bm \sigma}_1\cdot{\hat r}\ {\bm \sigma}_2\cdot{\hat r}
 - {\bm \sigma}_1\cdot {\bm \sigma}_2.
\ee
The radial functions $u_0(r)$ and $u(r)$ are normalized
so that their large-$r$ asymptotic functions $\phi_0(r)$ and $\phi(r)$
have $\phi_0(0)=\phi(0)=1$. We have
\bea
\lim_{r \rightarrow \infty} u_0(r) &=& \phi_0(r)
  \equiv \frac{\sin(p r + \delta_s)}{\sin(\delta_s)}
  \simeq 1 - \frac{r}{a_s},
\\
\lim_{r \rightarrow \infty} u(r) &=& \phi(r)\equiv \e^{-\gamma r},
\nonumber \\
\lim_{r \rightarrow \infty} w(r) &=& \eta\, \e^{-\gamma r} \left(
  1 + \frac{3}{\gamma r} + \frac{3}{(\gamma r)^2}\right)
\eea
where $\eta$ is the $D/S$ ratio,
$\gamma= \sqrt{B_d \mN} \simeq 45.702\ \mbox{MeV} \simeq
(4.3177\ \mbox{fm})^{-1}$,
$B_d\simeq 2.2246$~MeV is the deuteron binding energy, $a_s\simeq -23.749\
\mbox{fm}$ is the singlet $np$ scattering length
and $\delta_s$ is the ${}^1 S_0$ $np$ phase shift.
In terms of the M1 transition matrix $M$ defined by
\be
\frac{e}{2 m_p} M
\equiv
    \int\!\! d^3{\bm r}\,\, \Psi_{np}^\dagger({\bm r})
    \left[{\bm \mu}({\bm r})\right]^z \Psi_d ({\bm r}; 0),
\ee
the cross section can be written as
\be
\sigma = G\, M^2,\ \ \ \ \
G\equiv \frac{e^2 \omega^3 \, A_s^2\, a_s^2}{4\,v_n \,m_p^2}
\label{cross2}
\ee
where $v_n$ is the neutron velocity,
$\omega\simeq B_d$ is the energy of the emitted photon,
and $A_s \simeq 0.8846\ \mbox{fm}^{-1/2}$
is the normalization factor of the deuteron wavefunction,
\be
\frac{1}{A_s^2} \equiv
\int\!\! d^3{\bm r}\,\,
  \Psi_d^\dagger({\bm r}) \Psi_d ({\bm r}; 0)
= \int_0^\infty\!\! dr\, \left[ \,u^2(r) + w^2(r)\right].
\ee

Now the transition matrix $M_{\rm 1B}$
for the one-body magnetic moment operator (\ref{mu1B}) is
\be
M_{\rm 1B} = \mu_V\,\int_0^\infty\!\! dr\, u_0(r) u(r).
\ee
Austern showed in 1953 \cite{aust} that $M_{\rm 1B}$ can be
estimated accurately in an effective-range expansion
in terms of low energy parameters.
The leading contribution is
\be
M_0 \equiv \mu_V\, \int_0^\infty\!\! dr\, \phi_0(r) \phi(r)
= \frac{\mu_V}{\gamma}\, \left(1 - \frac{1}{\gamma\,a_s}\right).
\ee
The next order contribution in the effective range expansion
can also be represented \cite{aust}
by scattering lengths, effective ranges and
the $D$-wave probability of deuteron.
The impulse approximation obtained with this method comes out to be
$\sigma_{imp}(\mbox{Austern}) = (303 \pm 4)\ \mbox{mb}$.
As we will explain in the following subsection,
one can obtain a considerably more accurate value  with the Argonne $v_{18}$
potential \cite{v18},
\be
\sigma_{imp} = 305.6\ \mbox{mb}.\label{npimp}
\ee
The discrepancy between the impulse (\ref{npimp}) (or rather the Austern
result) and the experiment (\ref{npexp}) had been a long-standing puzzle
in nuclear physics until Riska and Brown \cite{riskabrown}
supplied in 1972 the missing 10~\%
contribution in terms of meson-exchange currents.
In the following subsection, we will do an accurate one-loop
chiral perturbation calculation, using the Argonne $v_{18}$ potential.
\subsection{Exchange current contribution}
\indent

In calculating the matrix element of the two-body currents $M_{\rm 2B}$,
it is convenient to perform the angular integration first
and evaluate the spin-isospin matrix elements for various
spin-isospin operators:
\bea
\langle {\hat T}_{S,T}^{(\odot)}\rangle_{\rm SS}
&=&
\int\!\frac{d\Omega}{4\pi}\,
\left(\chi_{00}\, \zeta_{10}\right)^\dagger
\left[ {\hat T}_{S,T}^{(\odot)}\right]^z
\chi_{1 0}\, \zeta_{00} ,
\nonumber \\
\langle {\hat T}_{S,T}^{(\odot)}\rangle_{\rm SD}
&=&
\int\!\frac{d\Omega}{4\pi}\,
\left(\chi_{00}\, \zeta_{10}\right)^\dagger
\left[ {\hat T}_{S,T}^{(\odot)}\right]^z
\,S_{12}(\hat{r})
\chi_{1 0}\, \zeta_{00} .
\eea
It is straightforward to show that
\be
\langle \TminusS \rangle_{\rm SS}
= \langle \TtimesS\rangle_{\rm SS} = 4,
\ \ \ \ \ \
\langle \TminusT \rangle_{\rm SD}
= \langle \TtimesT\rangle_{\rm SD} = - \frac{16}{3}
\label{ang:ave}\ee
and that all others are equal to zero.
The resulting two-body transition matrix elements are
\be
M_{\rm 2B}
= M_{\rm 2B}^{\rm SS} + M_{\rm 2B}^{\rm SD}
\ee
with
\bea
M_{\rm 2B}^{\rm SS} &=&
  \frac{\gA^2\,m_p\,\mpi}{\fpi^2}\,
\int_0^\infty\! \! dr\,u_0(r)\,u(r)\,
\left[f_S^{(-)}(x_\pi) + f_S^{(\times)}(x_\pi)\right] ,
\nonumber \\
M_{\rm 2B}^{\rm SD} &=&
-\frac{\sqrt{2} \gA^2\,m_p\,\mpi}{3 \fpi^2}\,
\int_0^\infty\! \! dr\,u_0(r)\,w(r)\,
\left[f_T^{(-)}(x_\pi) + f_T^{(\times)}(x_\pi)\right]
\eea
where the functions $f_{S,T}^{(-,\times)}(x_\pi)$ are
($x_\pi\equiv \mpi\,r$)
\bea
f_S^{(-)}(x_\pi) &=&
  - \frac{{\bar c}_\omega + {\bar c}_\Delta}{3}\, y_0(x_\pi)
  + \frac{\mpi^2}{48 \pi^2 \fpi^2}
    \, x_\pi\frac{d}{dx_\pi} k_0(x_\pi)
  + \frac{4 \fpi^2}{\gA^2\mpi} G_A^R \delta({\bm r}),
  \nonumber \\
f_S^{(\times)}(x_\pi) &=&
  \frac{{\bar c}_\Delta}{3} y_0(x_\pi) +
   \left[\frac{2}{3} y_1(x_\pi)- y_0(x_\pi) \right]
  \nonumber\\
  &+&
    \frac{\gA^2\,\mpi^2}{48 \pi^2 \fpi^2}
    \left[\frac12 k_0(x_\pi) - k_1(x_\pi)
    + x_\pi \frac{d}{d x_\pi} k_0(x_\pi)\right]
  + \frac{4 \fpi^2}{\gA^2\mpi} E_T^{V,R} \delta({\bm r}),
  \nonumber \\
f_T^{(-)}(x_\pi) &=&
  -({\bar c}_\omega + {\bar c}_\Delta)\,y_2(x_\pi)
  - \frac{\mpi^2}{32 \pi^2 \fpi^2}
    \, x_\pi\frac{d}{dx_\pi} k_0(x_\pi),
  \nonumber \\
f_T^{(\times)}(x_\pi) &=&
  - \frac{{\bar c}_\Delta}{2} \, y_2(x_\pi)
  + \frac{\gA^2\,\mpi^2}{32 \pi^2 \fpi^2}
  \left[2 k_{\rm T}(x_\pi) - x_\pi \frac{d}{d x_\pi} k_0(x_\pi)
  \right].
\eea

\subsection{The $\Delta$ degree of freedom}
\label{delta:S}
\indent

So far, we have not considered the $\Delta$ degrees of freedom explicitly.
It figured indirectly only as a
resonance that saturates one of the counter terms.
The role of the $\Delta$ is subtle because of the non-commutativity of
the large-$N_c$ limit and the chiral limit. The matter is simple
if the mass gap,
$\delta m= m_\Delta - \mN$,
is either small enough to be regarded as $\O(Q)$
or large enough to be regarded as $\O(\Lambda_\chi)$.
In the large-$N_c$ limit, $\delta m = \O(N_c^{-1})$ is to be taken to be
smaller than $m_\pi=\O(N_c^0)$.
On the other hand, in the chiral limit,
it is to be taken to be bigger than $m_\pi$ since it remains finite
when $m_\pi$ goes to zero. In reality, neither is a good approximation.
For the process we are considering, we believe that both approaches, one
with the $\delta m$ treated as small (referred to as ``type-I")
and the other with the $\delta m$ treated as heavy (referred to as
``type-II") could be employed equally well.

In the type-I approach, we should treat the nucleon and the $\Delta$ on the
same footing. Then the counting rule should be modified to
\be
\nu^{\rm I} = 4 - 2 C  - \left(\frac{E_N + E_\Delta}{2}
+  E_H + E_E \right) + 2 L
 + \sum_i {\bar \nu}^{\rm I}_i,\ \ \ \ \
{\bar \nu}^{\rm I}_i \equiv d_i + \frac{n_i + n_i^*}{2} + h_i  + e_i - 2
\label{countingI}\ee
where $E_\Delta$ is the number of external $\Delta$ lines and
$n_i^*$ the number of $\Delta$ lines attached to the $i$th vertex.
In this approach, we cannot integrate out the $\Delta$ degree of
freedom without introducing non-locality in the effective Lagrangian.
This also means that the $\Delta$ must be included in the ``reducible
graphs" or in terms of Schr\"odinger picture, in the wavefunctions.
This procedure has often been used in nuclear physics calculations for
processes that involve the excitation of the $\Delta$ resonance. Processes
involving large energy transfer are better treated in this manner.

In the type-II approach, we should be able to treat $\delta m$ as heavy,
say, $\delta m = \O(Q^0)$. In this case, we cannot
attach any $Q$-factor to the propagator,
since $1/ (\delta m - v\cdot k) \sim 1/\delta m = \O(Q^0)$. The counting
rule in this case is
\be
\nu^{\rm II} = 4 - 2 C
  - \left(\frac{E_N}{2}  + E_\Delta + E_H + E_E \right) + 2 L
 + \sum_i {\bar \nu}^{\rm II}_i,\ \ \ \ \
{\bar \nu}^{\rm II}_i \equiv d_i + \frac{n_i}{2} + n_i^* + h_i  + e_i - 2 .
\label{countingII}\ee
If this applies, we can freely integrate out the $\Delta$ degree of freedom
as well as other massive degrees of freedom. This will be the case for
the process we are considering. As shown recently
by Mallik \cite{mallik}, this approach is equivalent to the type-I approach
in certain kinematics (such as in our case).
It appears that the type-I approach is more general in the sense that
it can be applied to a wide range of kinematical conditions whereas
the type II is restricted.
For our case, the type-II approach is much simpler.
This can be seen as follows.
Since the energy carried by the exchanged pion (since the photon is
thermal) is small compared with the $N\Delta$ mass difference,
the $\Delta$ propagator in tree graphs may be safely
expanded as
\be
\frac{1}{\delta m - v\cdot k}
= \frac{1}{\delta m} + \frac{v\cdot k}{(\delta m)^2} + \cdots,\label{expansion}
\ee
where $v\cdot k$ corresponds to the energy of the pion or the photon.
In our one-loop calculation, only the first term is retained.
How good is this approximation? In general,
the $\Delta$ propagator in the two-pion-exchange
graphs may carry an arbitrary loop momentum. The only constraint
we have is that the loop momentum be of order of the characteristic
momentum, $Q$.
In our process, however, the characteristic momentum involved is
$\gamma=\sqrt{B_d\,\mN}\simeq 45$~MeV.
Therefore the error involved in truncating the series (\ref{expansion})
is at most of order $\gamma/(\delta m) \simeq 0.15$ in the
two-pion-exchange contributions.
Since the overall two-pion-exchange contribution to the thermal neutron
process is very small, as we shall find below, this is a safe approximation.

\subsection{Numerical results}
\indent

To calculate the capture cross section, we need accurate two-nucleon
wavefunctions.
This is because we need to have an accurate prediction for
${}^1S_0$ $np$ scattering length, $a_s$, to which
the cross section is directly proportional.
(There is also an implicit $a_s$-dependence through the wavefunction.)
A majority of nuclear potentials fail to reproduce
the experimental value $a_s^{exp}= -23.749 \pm 0.008$~fm,
because they do not fully incorporate the {\em charge-independence-breaking}
(CIB) effect of the potential.
It is not enough to incorporate the Coulomb interactions
and the mass difference between the charged and neutral pions. It is also
necessary to incorporate the CIB effects in intermediate- and short-range
potentials. Potentials which do not account for such CIB effects
predict typically $\sim a_s \sim -17$~fm.
For example, Reid's hard core potential \cite{reid}
predicts $a_s(\mbox{Reid})= -16.7$~fm,
giving a poor impulse approximation result for the capture rate,
$\sigma_{\rm imp}(\mbox{Reid})= 173.7$~mb to be compared with what is
needed, $\sim 306$ mb (see (\ref{argonne})). A similar result is obtained
with the Reid soft core potential, the Paris potential etc.
An important point to note is that the ratios $\delta$
of the matrix elements of the exchange currents
to the impulse current are remarkably insensitive to the potentials,
as we shall discuss later.

We take the Argonne
$v_{18}$ potential recently constructed by Wiringa, Stoks and Schiavilla
\cite{v18}. This potential is fit to 1787 $pp$ and 2514 $np$ scattering
data in the range 0--350 MeV with an excellent $\chi^2$ of 1.09 and gives
the deuteron properties -- the asymptotic S-state normalization, $A_S$,
the $D/S$ ratio, $\eta$, and the deuteron radius, $d_d$ -- close to the
experimental values.
Electromagnetic properties also come out
well, modulo exchange-current and relativistic corrections.
It predicts $a_s^{th}= -23.732$~fm in excellent agreement
with the experimental value $a_s^{{exp}}=-23.749\pm 0.008$~fm.
The single-particle matrix element with this potential gives the impulse
approximation cross section (given by ${\bm \mu}_{tree}$ of
eq.(\ref{mutotal}))
\be
\sigma_{imp} =305.6\ \mbox{mb},\label{argonne}
\ee
\noindent $\sim$9.6\% less than the experimental value
$\sigma_{exp}=334.2\pm 0.5$ mb.

\begin{figure}[tbp]
\centerline{\epsfig{file=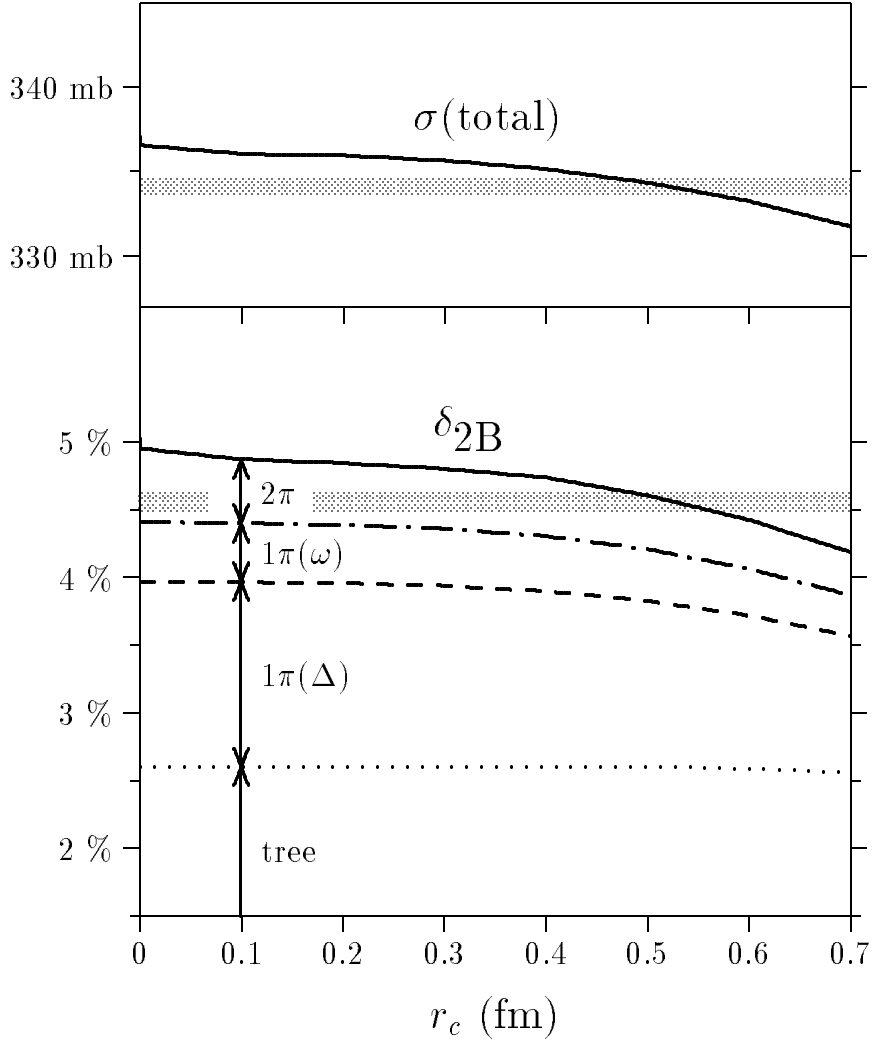}}
\caption{\protect \small
Total capture cross section $\sigma_{\rm cap}$ (top) and $\delta$'s (bottom)
vs. the cut-off $r_c$. The solid line represents the total
contributions and the experimental values are given by the shaded band
indicating the error bar.
The dotted line gives $\delta_{\rm tree}$, the dashed line
$\delta_{\rm tree} + \delta_{1\pi}^\Delta$, the dot-dashed line
$\delta_{\rm tree} + \delta_{1\pi}= \delta_{\rm tree} + \delta_{1\pi}^\Delta
+ \delta_{1\pi}^\omega$ and the solid line the total ratio, $\delta_{\rm 2B}$.}
\label{dataII}
\end{figure}

\begin{table}[tbp]
\begin{center}
\begin{tabular}{|c|c|c|c|c|c|c|c||c|}\hline
$r_c (\mbox{fm})$ &
   0.1 & 0.2 & 0.3 & 0.4 & 0.5 & 0.6 & 0.7
   & Reid \\
  \hline \hline
$\sigma$(mb) &
  336.1 & 336.0 & 335.7 & 335.2 & 334.4 & 333.3 & 331.8 & 188.8 \\ \hline
\hline
$\delta_{tree}$ (\%) &
  2.608 & 2.608 & 2.609 & 2.609 & 2.605 & 2.592 & 2.563 & 2.577 \\ \hline
$\delta_{1\pi}^\omega$ (\%) &
  0.435 & 0.431 & 0.423 & 0.408 & 0.382 & 0.346 & 0.301 & 0.367 \\ \hline
$\delta_{1\pi}^\Delta$ (\%) &
  1.364 & 1.355 & 1.334 & 1.294 & 1.228 & 1.132 & 1.011 & 1.168 \\ \hline
$\delta_{1\pi}$ (\%) &
  1.799 & 1.786 & 1.757 & 1.702 & 1.610 & 1.477 & 1.311 & 1.535 \\ \hline
$\delta_{2\pi}$ (\%) &
  0.471 & 0.457 & 0.443 & 0.424 & 0.397 & 0.362 & 0.319 & 0.319 \\ \hline
$\delta_{\rm 2B}$ (\%) &
  4.878 & 4.851 & 4.809 & 4.745 & 4.612 & 4.431 & 4.193 & 4.431 \\ \hline
\end{tabular}
\caption{The total cross section $\sigma$
and the $\delta$'s calculated
with the Argonne $v_{18}$ potential. The $\delta$'s are defined
by $\delta_x= M_x / M_{\rm 1B} \times 100$~\% for a given  $x$,
$\delta_{1\pi}= \delta_{1\pi}^\omega + \delta_{1\pi}^\Delta$,
and $\delta_{\rm 2B}=\delta_{\rm tree} + \delta_{1\pi} + \delta_{2\pi}$.
The last column shows the results with the Reid hard-core
potential.
\label{ArgonneI}}
\end{center}
\end{table}

\begin{table}[tbp]
\begin{center}
\begin{tabular}{|c|c|c|c|}\hline
& $SS$ & $SD$ & Sum ($SS + SD$) \\
\hline\hline
$\delta_{tree}$ (\%) & 1.534 & 1.071 & 2.605 \\ \hline
$\delta_{1\pi}^\omega$ (\%) & $-0.119$ & 0.501 & 0.382 \\ \hline
$\delta_{1\pi}^\Delta$ (\%) & 0 & 1.228 & 1.228  \\ \hline
$\delta_{1\pi}$ (\%) & $-0.119$ & 1.729 & 1.610 \\ \hline
$\delta_{2\pi}$ (\%) & 0.449 & $-0.052$ & 0.397 \\ \hline
$\delta_{\rm 2B}$ (\%) & 1.864 & 2.748 & 4.612 \\ \hline
\end{tabular}
\caption{The $\delta$'s with $r_c= 0.5$~fm
calculated with the Argonne $v_{18}$ potential.
\label{ArgonneII}}
\end{center}
\end{table}

In computing the matrix elements of the two-body operators, we need
to take into account short-range correlations in the wavefunctions.
Short-range correlations must involve nucleon interactions at a length
scale less than the scale given by chiral symmetry breaking and cannot
be accounted for by low-order chiral perturbation expansions. In calculating
Feynman diagrams of ChPT, the cut-off of the theory is incorporated in
the regularization. Therefore nuclear wavefunctions do not incorporate
the effect of the degrees of freedom that are
cut off from the theory and the short-range
correlation that is represented by a hard-core radius is missing.
We do not know how to implement this effect in a way consistent with the
strategy of ChPT. In this paper as in \cite{pmr1}, we account for the
short-range effect simply
by multiplying the integrand of the radial integral by $\theta (r-r_c)$,
$$\int_0^\infty\!\! dr \rightarrow
\int_{r_c}^\infty\!\! dr\,.
$$
This procedure is justified by the insensitivity of the result to the
hard-core size.

We use the physical values for masses and constants that appear in the
theory. There are no unknown parameters other than
the short-range cutoff parameter $r_c$.
The resulting total cross section is plotted
and compared with the experiment \cite{cox}
in Fig.~\ref{dataII} (top)
for a wide range of $r_c$, $0 < r_c \lsim 0.7$~fm.
Figure~\ref{dataII} (bottom) shows the contribution of each
term in terms of the ratios
$\delta_{tree}\equiv
M_{tree}/M_{\rm 1B}$, $\delta_{1\pi}\equiv M_{1\pi}/M_{\rm 1B}$,
$\delta_{2\pi}\equiv M_{2\pi}/M_{\rm 1B}$ and $\delta_{2B}$ which is
the sum.
For $\delta_{1\pi}$, we divided it into two part,
$\delta_{1\pi}^\omega$ and $\delta_{1\pi}^\Delta$.
The ratio corresponding to the
``generalized tree" contributions is
$\delta_{\rm GTO}= \delta_{tree} + \delta_{1\pi}$.
These ratios are plotted in Figure~\ref{dataII} (bottom).
We also list the total cross section
and the ratios in Table~\ref{ArgonneI} and \ref{ArgonneII}.
In Table~\ref{ArgonneI}, the results are listed for
varying $r_c$'s. The last column shows
the results with Reid's hard-core potential \cite{reid}.
{\it Notice that, although the cross section is very sensitive
to the potential, the ratios are essentially the same.} This confirms that
the ratios are highly model-independent, a conclusion reached
previously.\footnote{
What this means as far as the $np$ capture is concerned is that
all that is needed for an agreement with the experimental capture rate
is to fix parameters of finite-order
chiral perturbation theory so as to fit
the singlet scattering length $a_s$, since the ratios of the matrix elements
are insensitive to nuclear potentials.
It thus appears that such a refined potential as the Argonne $v_{18}$ potential
is not really necessary for a satisfactory result.
We are grateful to Jim Friar for emphasizing this point.}
We should also point out that the $r_c$
which gives approximately the same result
with Reid's potential is $r_c\simeq 0.6$~fm to be compared with
the hard-core radius of the Reid potential, $0.55$~fm.
In Table~\ref{ArgonneII}, we decompose the ratios
into the $SS$ and $SD$ channels (where $SS$ ($SD$) denotes the
contribution to the $S$-wave ($D$-wave) of the
deuteron wavefunction)
for a fixed $r_c$, $r_c=0.5$~fm.
The intrinsic uncertainty associated
with short-distance physics notwithstanding, the theoretical
prediction
\be
\sigma_{th}=334\pm 3\ \mbox{mb}
\ee
\noindent (where the theoretical
error bar represents only the dependence on the hard-core cut-off)
is in remarkable agreement with the experiment, say, within less than 1\% !
We have checked that the isoscalar moment contribution to the process
is totally negligible.\footnote{A rough estimate using Figure 7(d) as the
isoscalar exchange-current operator shows that
the isoscalar contribution to the cross section is suppressed relative
to the isovector one by  a factor of more than $10^5$.}

\subsection{The chiral-filter phenomenon}
\indent

One of the principal results in our calculation is that the
``generalized tree" contributions dominate to the \NNL\ order, with only
a small correction
(less than 0.6 \% of the single-particle matrix element) coming from the
genuine one-loop correction. This agrees with the ``chiral filter"
mechanism seen in the axial-charge transitions \cite{pmr1} and confirms
the conjecture made in \cite{kdr}. We should make a few additional
remarks on this point.

It is interesting to compare the
M1 transition (thermal neutron capture)
with the axial-charge transitions \cite{pmr1}.
In fact, the meaning of ``chiral filter"
is slightly different in the two :
In axial-charge transitions, the chiral filter mechanism is reflected
in the smallness of the loop contribution
compared with the tree contribution whereas in M1 processes, it is in the
suppression of the two-pion exchange (one-loop) term compared with
the total one-pion exchange (generalized tree) contribution.
If we denote the operator by $\M^A$ ($\M^V$)
and its matrix element by $M^A$ ($M^V$)
for axial-charge (M1) transitions, then the statement is
\bea
M_{loop}^A &=& M_{1\pi}^A + M_{2\pi}^A \ll M_{tree}^A,
\\
M_{2\pi}^V &\ll& M_{GTO}^V = M_{tree}^V + M_{1\pi}^V.
\eea
These two observations can be unified if one accepts that
what chiral filter conjecture really
says is that the long-range operator (characterized by
its tail $\e^{-m_\pi r}$)
dominates whenever it is not suppressed by kinematical or symmetry reasons.
In the axial-charge transition case,
as was discussed in \cite{pmr1},
the $\M_{1\pi}^A$ can be decomposed into two parts,
long-ranged and short-ranged as
\be
\M_{1\pi}^A = \delta_{soft} \M_{tree}^A + \M_{1\pi;short}^A
\ee
with
\be
\delta_{soft} = c_3^R \frac{m_\pi^2}{f_\pi^2}
+ \frac{m_\pi^2}{16\pi^2 f_\pi^2}\left[ \frac{1+3\gA^2}{2}
\left(2-\frac{\pi}{\sqrt{3}}\right) - (1+2\gA^2) \left(
\frac{17}{9} - \frac{\pi}{\sqrt{3}} \right)\right]
\simeq 0.051
\ee
where
$\M_{1\pi;short}^A$ represents the short-range contribution (shorter-ranged
compared to one-pion-exchange) and
the counter term $c_3^R$ is given by the charge radius of the proton.
In \cite{pmr1}, we showed that $\delta_{soft}$ (or $c_3^R$) could be understood
in terms of the $\rho$-meson degree of freedom.
Since two-pion-exchange contributions are short-ranged,
we can clearly divide the exchange axial-vector current
into a long-ranged component and a residual short-ranged part:
\bea
\M_{long}^A &=& (1 + \delta_{soft}) \M_{tree}^A,
\nonumber \\
\M_{short}^A &=& \M_{1\pi; short}^A + \M_{2\pi}^A.
\eea
Put in this form, the chiral filter effect is even more transparent
and striking
\be
\left[\frac{M_{short}}{M_{long}}\right]^A \simeq
(0.03 - 0.05)
\ee
with the matter density ranging $0 < \rho < \rho_0$.
In the M1 transition in question, the corresponding long- and short-ranged
operators are
\bea
\M_{long}^V &=& \M_{tree}^V + \M_{1\pi;\omega}^V + \M_{1\pi;\Delta}^V
= \M_{GTO},
\nonumber \\
\M_{short}^V &=& \M_{2\pi}^V.
\eea
We see that in this extended form the chiral filter works equally well
for both processes.

\section{The BR Scaling}
\indent

There is very little understanding of
the meaning of the short-distance correlation cut-off
$r_c$ in the context of ChPT.  As discussed in \cite{pmr1},
the loop terms contain zero-range operators in coordinate space. In addition,
four-Fermi counter terms in the chiral Lagrangian with
unknown constants are of contact interaction. At higher chiral order,
increasingly shorter-ranged operators would enter together
with the zero-ranged ones.
Now if we were able to compute nuclear interactions to all orders in
chiral perturbation theory,
the delta functions in the current would be naturally regularized
and would cause no problem. Such a calculation  of course is an
impossible feat.
The practical application of
chiral perturbation theory is limited to low orders in the chiral expansion,
so a cut-off would be needed to screen the interactions
shorter-ranged than accessible by the chiral expansion adopted. This means
that {\it all} interactions of range shorter than the inverse chiral
scale $\sim 1\ {\mbox{GeV}}^{-1}$ should not contribute as additional
operators.
Clearly such a calculation would be meaningful only if the dependence on
the cut-off were weak. Our calculation here meets that criterion.
In the present work, we find that
the $r_c$-dependence is trivially small over wide range, $0 < r_c \lsim
0.5$~fm,
and the theoretical prediction agrees exactly with the experiment
at $r_c\simeq 0.5$~fm.
The $r_c$-independence and the success can be traced to the fact that
there is no relevant low-energy resonances which contribute to the counter
terms
associated with four-Fermi interactions and the fact that
the short-range correlation effect is automatically included in the
$np$ and deuteron wavefunctions.

There appear subtleties, however, when nuclear density goes up, as in the
case of the axial-charge transitions in heavy nuclei studied elsewhere.
As nuclear density increases, some of the massive degrees of freedom that
have been integrated out
could come down in energy as suggested by Brown and Rho\cite{br91,br95}.
If the mass comes down below the relevant chiral scale, roughly equivalent to
$r_c^{-1}$, then the short-range correlation can no longer suppress such
a contribution. We shall now suggest that this is what happens with a scalar
``dilaton" field.

Consider the scalar field $\chi$ associated with the trace anomaly of
QCD exploited in \cite{br91,br95}. In matter-free space ($\rho\approx 0$),
the scalar field, associated with scalar glueball, is massive with
a mass $m_\chi\sim 2\ {\mbox{GeV}} > m_\rho$, so may be integrated out.
A simple calculation shows that this leads to an
$O(Q^3)$ four-Fermi counter-term Lagrangian of the form
\be
\Delta {\cal L}=
{\alpha} (\Bbar B)
\left(\Bbar \Gamma_{1/m} B\right)
\label{sigmaL}\ee
where $\alpha$ is a dimension--4 constant
and $\Gamma_{1/m}$ is the operator which appears
in $1/\mN$ expansion, eq.(\ref{Gamma1m}),
$$ \Gamma_{1/m} = - D^2 + (v\cdot D)^2 + S^{\mu\nu} \Gamma_{\mu\nu}
+ \gA^2 (v\cdot \Delta)^2 + 2\gA \left\{v\cdot \Delta, S\cdot D\right\}
+ \cdots . $$
One can see from dimensional counting that this Lagrangian is highly
``irrelevant" and hence would be negligible. However the situation
can be drastically different in medium. As discussed in \cite{br95},
a low-lying excitation with the quantum number of scalar meson exists
in dense medium. Such a ``dilaton" is supposed to play
an important role in nuclear physics as manifested in Walecka's theory with
a low-lying scalar meson and also required by a ``mended symmetry"
emerging at a higher energy-momentum scale \cite{beane}. What this means
is that instead of the zero-range interaction coming from the four-Fermi
interaction, we would have a finite-range interaction with the range determined
by the scalar mass in medium, $m^\star_\sigma$ where $\sigma$ is the
scalar corresponding to the dilaton.
We can gain a simple idea by noting that the Lagrangian (\ref{sigmaL})
can be obtained from the ``$1/\mN$" Lagrangian (\ref{Lag1})
by replacing the nucleon mass by its effective operator,
\be
\mN \rightarrow \mN^* = \mN \left(1 + \kappa \frac{\Bbar B}{\rho_0}
\right)^{-1}
\ee
where $\rho_0$ is the normal nuclear matter density
and $\kappa$ is a constant.
If one assumes that $\kappa$ is saturated by the $\sigma$ meson effective
in medium \cite{br95},
then one finds
\be
\kappa = \frac{g_\sigma^2\,\rho_0}{\mN\,m_\sigma^2}
\label{sigcont}
\ee
where $g_\sigma$ is the $\sigma NN$ coupling and $m_\sigma$ the $\sigma$
mass. In mean-field approximation, (\ref{sigmaL}) and (\ref{sigcont})
give, for
$m_\sigma\approx 600 {\mbox{MeV}}$ and $g_\sigma\approx 10$,
a substantial enhancement to the one-body axial charge operator
as well as to the one-body isovector magnetic moment operator:
\bea
\frac{\delta_\sigma A^0}{A^0} = \frac{\delta_\sigma {\bm \mu}}{{\bm \mu}}
= \kappa\,\frac{\rho}{\rho_0} \simeq 0.4\ \frac{\rho}{\rho_0}.
\eea
This is basically the enhancement in the axial charge in heavy
nuclei as found in phenomenological models with
a low-lying scalar meson by Towner \cite{ian} and by Kirchbach {\it et al}
\ \cite{kirsch}. It is equivalent to
the Brown-Rho scaling $f_\pi^\star/f_\pi\approx m_N^\star/m_N$
as discussed in \cite{kr}.

\section{Conclusion}
\indent

This paper provides a chiral symmetry interpretation of the long-standing
exchange-current result of Riska and Brown. It supports the chiral-filter
argument verified in the axial-charge transitions and offers yet
another evidence that chiral Lagrangians figure importantly in nuclear physics.

Here as well as in \cite{pmr1}, ChPT is developed in the long wavelength
regime where the expansion makes sense. The question remains as to at what
point the expansion of ChPT breaks down. In particular, the meaning
of short-range correlation remains to be clarified, an issue which cannot be
ignored in any nuclear process. An intriguing question in this connection
is posed by the observation that the generalized one-pion exchange process
dominates even in $e + d\rightarrow n + p +e^\prime$ at a large
momentum transfer\cite{fm}. The only way one can understand this phenomenon
would be that the chiral filter takes place even to large momentum
transfers. It would be interesting to verify this in a systematic
chiral perturbation calculation.

Another intriguing problem is the role of many-body interactions from
chiral symmetry point of view. We know that for processes involving
small energy transfers, three-body and other higher-body currents
are suppressed. But this will not be the case for non-negligible
energy transfers. There multi-body currents will certainly become
important. Furthermore as nuclear density increases (i.e., heavy nuclei
or nuclear matter), particle masses can
drop below the characteristic chiral mass scale and introduce density-dependent
renormalization in the parameters of the chiral Lagrangian. We have illustrated
this with BR scaling \cite{br91} but the basic issue can be more
general in the sense that it involves relevant and irrelevant degrees of
freedom as the length scale is varied. This problem needs further study.

\subsection*{Acknowledgments}
\indent

We thank the Institute for Nuclear Theory at the
University of Washington for its hospitality
and the Department of Energy for partial support
during the completion of this work.
The work of TSP and DPM is supported in part by the Korea
Science and Engineering Foundation through the
Center for Theoretical Physics of Seoul National University
and in part by
the Korea Ministry of Education under the grant No.BSRI-94-2418.

\section*{Appendices}
\renewcommand{\thesubsection}{\Alph{subsection}}
\renewcommand{\theequation}{\Alph{subsection}.\arabic{equation}}
\setcounter{subsection}{0}

\subsection{Renormalization of the ${\cal V}\pi NN$ vertex}
\label{VpiNN:A}
\setcounter{equation}{0}
\indent

Consider the ${\cal V}\pi NN$ vertex $\GpiV(k,q)$ defined by the kinematics
$$N(\mN v) \rightarrow N(\mN v-k-q) + \pi_b(q) + {\cal V}_a(k).$$
Restricting ourselves to on-shell nucleons ({\it i.e.}, $v\cdot(k+q)=0$) and
including the $1/ \mN$ corrections, the vertex has the form
\bea
\GpiV &=& - \epsilon_{abc}\tau_c\frac{\gA}{\fpi}\,S^\mu \nonumber \\
&+& \epsilon_{abc}\tau_c\left[-\frac{\gA}{\fpi^3}S^\mu y\,h_0^A(y)
+ \frac{\gA^3}{\fpi^3} v^\mu\frac{d-3}{4} q\cdot S \,h_4^A(y)
-\frac{\gA}{\fpi^3} y \,I_A^\mu(k)\right]
\nonumber \\
&+& i\delta_{ab}\left[ 2 \frac{\gA}{\fpi^3} S^\mu y \,h_0^S(y)
-\frac34 \frac{\gA^3}{\fpi^3} v^\mu \,q\cdot S \,h_4^S(y)
-2 \frac{\gA}{\fpi^3} y \,I_S^\mu(k) - 2 \frac{\gA^3}{\fpi^3}
q_\nu I^{\mu\nu}(k)\right]\nonumber \\
&+& \frac{v\cdot q}{2 \mN \fpi} \, S^\mu\,\left[
i \delta_{ab}\, (\gA + b_3) + \epsilon_{abc} \tau_c\, b_5 \right]
\nonumber \\
&+& \GpiVct
\label{piV}\eea
where
$\GpiVct$ are the contributions from the counter-term Lagrangian,
$y=v\cdot k$ and
$h_{0,4}^{A,S}(y) = \frac12 \left[h_{0,4}(y) \mp h_{0,4}(-y)\right]$
which were given explicitly in \cite{pmr1}.
The functions $I_{A,S}^\mu(k)= \frac12\left[I^\mu(k) \mp I^\mu(-k)\right]$
and $I^{\mu\nu}(k)$ are defined by
\bea
I^\mu(k) &=& \int_l \frac{S\cdot l\,(2l+k)^\mu}{v\cdot l\,
(l^2-M^2)\,\left[(l+k)^2-M^2\right]},
\nonumber \\
I^{\mu\nu}(k) &=& \int_l \frac{S\cdot (l+k)\,S^\nu\,S\cdot l\,(2l+k)^\mu}{
v\cdot(l+k)\,v\cdot l\, (l^2-M^2)\,\left[(l+k)^2-M^2\right]}
\eea
where
$$\int_l \equiv \frac{\mu^{4-d}}{i}\int\!\!
\frac{d^d l}{(2\pi)^d},$$
$d= 4 - 2 \epsilon$ is the spacetime dimension
and $\mu$ is the renormalization scale.
For small momentum, they can be simplified to
\bea
I^\mu(k) &=&
-\frac{\mpi}{8\pi}\,S^\mu
- \left( S^\mu\, v\cdot k + v^\mu\, S\cdot k\right)
\ldiv + \O(k^2), \nonumber \\
I^{\mu\nu}(k) &=& - \frac{\mpi}{32\pi} v^\mu\,S^\nu - \frac{i}{4}
\epsilon^{\mu\nu\alpha\beta} v_\alpha k_\beta \ldiv + \O(k^2)
\eea
where $\O(k^2)$ denotes terms
equal to, or higher than, second order in
$k^\mu$ and $\epsilon_{0123}=1$.
With $v\cdot k=0$, they become
\bea
I^\mu(k) &=& -v^\mu \, S\cdot k \, f_0(k^2)
\nonumber \\
&-& \frac{1}{16\pi} \int_0^1 dz
\left(2 S^\mu\sqrt{\mpi^2-z(1-z) k^2} + q^\mu\, S\cdot q\, \frac{z(2z-1)}{
\sqrt{\mpi^2-z(1-z)k^2}}\right),
\nonumber \\
I^{\mu\nu}(k) &=& - \frac{i}{4} \epsilon^{\mu\nu\alpha\beta} v_\alpha k_\beta
f_0(k^2)
\nonumber \\
&-& \frac{v^\mu}{8\pi} \left(\frac14 S^\nu + 2 S\cdot k\, S^\nu\,
S\cdot k \frac{\partial}{\partial k^2}\right)
\int_0^1 dz \sqrt{\mpi^2-z(1-z)k^2}
\eea
where $f_0$ is \cite{pmr1}
\bea
f_0(q^2) &=& \ldiv - \frac{1}{16\pi^2} K_0(q^2)
\label{f0}\eea
and the divergent quantity $\ldiv$ is defined in eq.(\ref{LL}).

All the divergences in eq.(\ref{piV}) can be removed by the counter-term
Lagrangian (\ref{counterterm}), that is, the terms with $c$'s,
whose contribution to the ${\cal V}\pi NN$ vertex is
\bea
\GpiVct &=&
\epsilon_{abc}\tau_c\left[
  -\frac{\gA}{2 \fpi^3}\,c_2\, q\cdot S\, v^\mu\,v\cdot k
  +\frac{\gA}{2\fpi^3} c_8 v\cdot q\left(v^\mu\, k\cdot S
  - S^\mu\,v\cdot k\right)\right]
\nonumber \\
&+&i\delta_{ab}\left[ i\frac{\gA^3}{2\fpi^3} c_7 \epsilon^{\mu\nu\alpha\beta}
  v_\alpha k_\beta q_\nu\right]
\nonumber \\
&-& \frac{i\,\gA}{4 \fpi^3}\,c_9\, \left[
  \Lambda_{ba} \,q_\beta \calP^{\beta\alpha}
    (k_\alpha\, S^\mu - g_\alpha^\mu\,k\cdot S)
 - \Lambda_{ab} \,(S^\mu\,k_\alpha - g^\mu_\alpha\,k\cdot S)
     \calP^{\alpha\beta} q_\beta \right]
\eea
with
\bea
c_2 &=& \frac{d-3}{3} \gA^2 \ldiv + c_2^R,\nonumber \\
c_7 &=& -\ldiv + c_7^R, \nonumber \\
c_8 &=& 2 \ldiv + c_8^R.
\eea
Here $c_i^R$ are finite, $\mu$-independent renormalized constants.
When $v\cdot k =v\cdot q = 0$ we have a simpler expression:
\bea
\GpiV(k,q) &=& - \epsilon_{abc}\tau_c \frac{\gA}{\fpi} S^\mu
- \delta_{ab} \frac{\gA^3}{2\fpi^3} \epsilon^{\mu\nu\alpha\beta}
v_\alpha k_\beta q_\nu \left[f_0(k^2)-f_0(0) + c_7^R\right]
\nonumber \\
&-& \frac{i\,\gA}{4 \fpi^3}\,c_9\, \left[
  \Lambda_{ba} \,q_\beta \calP^{\beta\alpha}
    (k_\alpha\, S^\mu - g_\alpha^\mu\,k\cdot S)
 - \Lambda_{ab} \,(S^\mu\,k_\alpha - g^\mu_\alpha\,k\cdot S)
     \calP^{\alpha\beta} q_\beta \right].\nonumber\\
\eea
Terms proportional to $v^\mu$ do not figure in our calculation and so
are omitted.
Note that the above equation is finite and $\mu$-independent and
contains two parameters, $c_7^R$ and $c_9$ which were fixed in our calculation
by resonance saturation.

\subsection{Renormalization of two-pion-exchange contributions}
\label{Jem:A}
\setcounter{equation}{0}
\indent

The two-body vector currents come from the two-pion exchange graphs
and one-pion exchange graphs with a four-Fermi contact vertex (see
Figure~\ref{TwoV}):
\bea
\Jem^\mu(a) &=&
   - (2 \tau_2 - i \tt)^z \frac{\gA^2}{4 \fpi^4} \, S_1^\mu\, S_1\cdot q_2\,
   f_0(q_2^2)
   \nonumber \\
&+& (2 \tau_2 + i \tt)^z\, \frac{\gA^2}{4 \fpi^4} \, S_1\cdot q_2 \, S_1^\mu\,
   f_0(q_2^2)+ (1\leftrightarrow 2),
   \nonumber \\
\Jem^\mu(b) &=& - (2 \tau_2 - i \tt)^z\, \frac{\gA^2}{2 \fpi^4} \, \int_l
   \frac{S_1\,\cdot(l-q_1)\, S_1\cdot l\, (2l+q_2-q_1)^\mu}{
   (l^2- \mpi^2)\,\left[(l-q_1)^2- \mpi^2\right] \left[(l+q_2)^2-\mpi^2\right]}
   \nonumber \\
&-&
   (2 \tau_2 + i \tt)^z\, \frac{\gA^2}{2 \fpi^4} \, \int_l
   \frac{S_1\cdot l\, S_1\,\cdot(l+q_1)\, (2l+q_1-q_2)^\mu}{
   (l^2- \mpi^2)\,\left[(l+q_1)^2- \mpi^2\right] \left[(l-q_2)^2-\mpi^2\right]}
   \nonumber \\ &&\ + \ (1\leftrightarrow 2), \nonumber \\
\Jem^\mu(c)
&=& - (2 \tau_1 + 2 \tau_2 - i \tt)^z\, \frac{\gA^4}{\fpi^4}\,
   S_1^\mu S_1^\alpha S_2^\beta S_2^\nu\, F_{\alpha\beta\nu}(q_2)
   \nonumber \\
&-& (2 \tau_1 + 2 \tau_2 - i \tt)^z\, \frac{\gA^4}{\fpi^4}\,
   S_1^\nu S_1^\alpha S_2^\beta S_2^\mu\, F_{\alpha\beta\nu}(-q_1)
   + (1\leftrightarrow 2), \nonumber \\
\Jem^\mu(d) &=&
   (2 \tau_1 + 2 \tau_2 - i \tt)^z\, \frac{\gA^4}{\fpi^4}
   \,\otimes
   \nonumber \\
   &&\ \int_l
   \frac{S_1\cdot(l-q_1)\, S_1\cdot l\, S_2\cdot l\, S_2\cdot(l+q_2)\,
   (2l+q_2-q_1)^\mu}{ (v\cdot l)^2\, (l^2-\mpi^2)\,\left[(l-q_1)^2-\mpi^2
    \right]\, \left[(l+q_2)^2-\mpi^2\right]}
    + (1\leftrightarrow 2), \nonumber \\
\Jem^\mu(e) &=&
   - 2i \epsilon^{3bc}\, \frac{\gA^2}{\fpi^2}\,\sum_A C_A \int_l
   \frac{S_1\cdot\left(l+\frac{k}{2}\right)\ \tau_1^b\Gamma_1^A\tau_1^c\
   S_1\cdot\left(l-\frac{k}{2}\right)\ \Gamma_2^A\ l^\mu}{ (v\cdot l)^2
   \left[\left(l-\frac{k}{2}\right)^2-\mpi^2\right]\,\left[\left(l+\frac{k}{2}
   \right)^2-\mpi^2\right]}
   \nonumber \\ &&\ + \ (1\leftrightarrow 2),
   \nonumber \\
\Jem^\mu(f) &=&
   -i\ttz\,\frac{1}{4 \fpi^4}\,
   \int_l \frac{(v\cdot l)^2\,(2l+q_2-q_1)^\mu}{(l^2-\mpi^2)\,
   \left[(l-q_1)^2-\mpi^2\right]\,\left[(l+q_2)^2-\mpi^2\right]},
   \nonumber \\
\Jem^\mu(g)
   &\propto& v\cdot S \, S^\mu \,=\, 0
\label{VmuTwoPi}\eea
where
\bea
F_{\alpha\beta\nu}(q)
&=& \int_l \frac{l_\alpha\, l_\beta\, (l+q)_\nu}{
   (v\cdot l)^2\, (l^2-\mpi^2)\,\left[(l+q_2)^2-\mpi^2\right]}
\nonumber \\
&=& \frac12 \left(g_{\alpha\nu} q_\beta + g_{\beta\nu} q_\alpha
- g_{\alpha\beta} q_\nu \right) f_0(q^2) + \frac{1}{16\pi^2} K_1(q^2)
\frac{q_\alpha q_\beta q_\nu}{-q^2} + (\mbox{$v$'s}).
\eea
Here and in what follows, ($v$'s) generically stands for terms proportional
to $v_\alpha,\ v_\beta$ or $v_\nu$.
The function $f_0(q^2)$ was defined in eq.(\ref{f0})
and the $K_1(q^2)$ was given in \cite{pmr1},
\bea
K_0(q^2) &=& \int_0^1\!\!dz\, \ln\left[ 1 - z(1-z) \frac{q^2}{m_\pi^2}
\right],
\nonumber \\
K_1(q^2) &=& \int_0^1\!\!dz\, \frac{ - z(1-z) q^2}{m_\pi^2  - z(1-z) q^2}.
\eea

Since for the M1 transitions, it is sufficient to retain only the  terms
linear in $k$, we define
\be
\Jem^{\mu,\nu} \equiv \left.
\frac{\partial \Jem^\mu}{\partial k_\nu}\right|_{k=0}\,.
\ee
A direct calculation leads to
\bea
\Jem^{\mu,\nu}(b) &=& - \Jem^{\nu,\mu}(a),
\nonumber \\
\Jem^{\mu,\nu}(d) &=& - \Jem^{\nu,\mu}(c) + \delta \Jem^{\mu,\nu}(d)
\eea
with
\be
\delta \Jem^{\mu,\nu}(d) = - (2\tau_1+2\tau_2-i\tt)^z\,
   \frac{\gA^4}{2 \fpi^4} \left(S_1^\mu S_1^\alpha S_2^\beta S_2^\nu
- S_1^\nu S_1^\alpha S_2^\beta S_2^\mu\right) F_{\alpha\beta}(q)
+ (1\leftrightarrow 2)
\ee
and
\bea
F_{\alpha\beta}(q) &=& \int_l \frac{l_\alpha \, l_\beta}{
(v\cdot l)^2\, (l^2-\mpi^2)\,\left[(l+q)^2-\mpi^2\right]}
\nonumber \\
&=& - g_{\alpha\beta} f_0(q^2) + \frac{q_\alpha q_\beta}{8\pi^2} \int_0^1 dz\,
\frac{z^2}{\mpi^2 - z(1-z) q^2} + (\mbox{$v$'s})
\eea
where ($v$'s) denotes terms proportional to $v_\alpha$ or $v_\beta$.

The divergencies appearing in the one-loop graphs, eq.(\ref{VmuTwoPi}),
are removed by the counter-term contribution coming from
the terms with $G$'s and $E_T^V$ in the counter-term Lagrangian
eq.(\ref{counterterm}).
Summing up all the two-pion-exchange and counter-term contributions,
we have
\bea
\Jem^{\mu,\nu}(2\pi)
&=& \left\{\frac{\gA^2}{4 \fpi^4} \frac{\del}{\del q_\nu}
\left(\tau_2^z\left[S_1^\mu,\ S_1\cdot q \right]\,{\bar f}_0(q^2)\right)
\right.
\nonumber \\
&+& \left.
(2\tau_1 + 2 \tau_2 -i\tt)^z \frac{\gA^4}{2 \fpi^4} \left(
\frac{\del}{\del q_\nu}
{\bar F}_1^\mu(S_1,S_2; \,q) - S_1^\mu S_1^\alpha S_2^\beta S_2^\nu
{\bar F}_{\alpha\beta}(q)\right)\right\}
\nonumber\\
&-&(\mu\leftrightarrow \nu)\ + (1\leftrightarrow 2)
\nonumber\\
&+&J_{\rm finite}^{\mu,\nu}(2\pi)
\label{vmn2loop}\eea
where
\bea
{\bar F}_1^\mu(S_1,S_2;\,q) &=&\left(S_1^\mu S_1^\alpha S_2^\beta S_2^\nu-
S_1^\nu S_1^\alpha S_2^\beta S_2^\mu\right)\ {\bar F}_{\alpha\beta\nu}(q),
\nonumber \\
{\bar F}_{\alpha\beta\nu}(q)
&=& -\frac{1}{32\pi^2} \left(g_{\alpha\nu} q_\beta + g_{\beta\nu} q_\alpha
- g_{\alpha\beta} q_\nu \right) K_0(q^2) + \frac{1}{16\pi^2} K_1(q^2)
\frac{q_\alpha q_\beta q_\nu}{-q^2} + (\mbox{$v$'s}),
\nonumber \\
{\bar F}_{\alpha\beta}(q)
&=& \frac{g_{\alpha \beta}}{16\pi^2} K_0(q^2)
+ \frac{q_\alpha q_\beta}{8\pi^2}
\int_0^1 dz\, \frac{z^2}{\mpi^2 - z(1-z) q^2} + (\mbox{$v$'s})
\eea
and
$J_{\rm finite}^{\mu,\nu}(2\pi)$ is the sum of the
divergent part of the loop graphs and the counter-term contribution,
\bea
J_{\rm finite}^{\mu,\nu}(2\pi) &=&
G_S^R\,(\tau_1+\tau_2)^z \left(S_1^{\mu\nu} + S_2^{\mu\nu} \right)
- G_A^R\,(\tau_1-\tau_2)^z \left(S_1^{\mu\nu} - S_2^{\mu\nu} \right)
\nonumber \\
&-& \frac{i}{2} \,E_T^{V,R}\,\ttz\,
\left(S_1^{\mu\alpha} S^\nu_{2\alpha} - S_1^{\nu\alpha} S^\mu_{2\alpha}\right).
\label{vmn2polect}\eea

\subsection{Resonance-exchange contributions}
\label{Res:A}
\setcounter{equation}{0}
\indent

In this appendix, we write down an effective chiral Lagrangian that contains
the $\Delta(1232)$ and vector mesons ($\rho^\mu$ and $\omega^\mu$)
in addition to $\pi$ and $N$ and apply it to the exchange currents.
Instead of writing a Lagrangian with all low-lying resonances in
full generality, we shall write only those terms relevant to nuclear
processes which can be compared with successful phenomenological model
Lagrangians employed in nuclear physics. (The axial field $a_1$
can be incorporated readily but we will not consider here.)

In heavy fermion formalism, including
the ``$1/\mN$" terms, the relevant Lagrangian is
\bea
{\cal L}^{\rm res}
  &=& \Bbar \left( i v\cdot \calD + 2 \gA \, S\cdot  i \Delta\right) B
+ \fpi^2 \Tr (  i \Delta_\mu\,  i \Delta^\mu)
+ \frac{\fpi^2}{4} \mpi^2 \Tr(\Sigma+\Sigma^\dagger)
\nonumber \\
&+& \frac{1}{2\mN} \Bbar \left( - \calD^2 + (v\cdot \calD)^2
- \gA^2 (v\cdot  i \Delta)^2 - 2i \gA \left\{v\cdot
 i \Delta , S\cdot \calD\right\} \right) B
\nonumber \\
&-& \frac{i}{2 \mN} \Bbar \left(
g_\rho \mu_{\rm V} [S^\mu,S^\nu] \rho_{\mu\nu}
+ \frac{g_\omega}{2} \mu_{\rm S}  [S^\mu,S^\nu] \omega_{\mu\nu}
\right) B
\nonumber \\
&+& m_\rho^2 \Tr\!\left(
\rho_\mu -\frac{1}{g_\rho} i \Gamma_\mu\right)^2
+ \frac{1}{2} m_\omega^2 \left(
\omega_\mu -\frac{2}{g_\omega} {\cal B}_\mu\right)^2
-\frac{1}{2} \Tr\, (\rho_{\mu\nu} \rho^{\mu\nu})
-\frac{1}{4} \omega_{\mu\nu} \omega^{\mu\nu}
\nonumber \\
&+&\L_\Delta + {\cal L}_{an}
\label{ResL}\eea
with
$\calD_\mu = \del_\mu - i g_\rho \rho_\mu - \frac{i}{2} g_\omega \omega_\mu$,
$\rho_{\mu\nu} = \del_\mu \rho_\nu - \del_\nu \rho_\mu
- i g_\rho \left[ \rho_\mu,\rho_\nu\right]$,
\be
{\cal L}_\Delta = {\bar T}_{\mu}
\left(\ddm -i v\cdot \calD \right) T^\mu
+ \left[{\bar T}^\mu \left(- {\cal C}\,  i \Delta_\mu
  - i \frac{g_\rho \muT}{2 \mN}\rho_{\mu\nu} S^\nu\right) B
  + \,\mbox{h.\,c.}\right]\label{DeltaL}
\ee
and
\be
\L_{an} = g_{\omega\rho\pi} \epsilon^{\mu\nu\alpha\beta}
{\vec \pi}\cdot\del_\mu {\vec \rho}_\nu \del_\alpha \omega_\beta
+ \cdots
\label{I6}\ee
In (\ref{DeltaL}),
$\ddm= m_\Delta-\mN \simeq 293$ MeV,
${\bar T} X T \equiv {\bar T}_{jkl} X^j_m T^{mkl}$ and
${\bar T} X B \equiv {\bar T}_{jkl} X^j_m B^k\, \epsilon^{3lm}$.
{}From the $\Delta$ decay width, we get ${\cal C}= -1.73$.
The value of $\muT$ has been determined \cite{jenkinsetal} to be
$\muT= -7.7\pm 0.5$.
The anomalous parity term (\ref{I6}) is the same as in \cite{fuji}
with\footnote{
In \cite{fuji}, Fujiwara {\it et al}\ considered anomalous parity terms in
$U(3)$ with $g_\omega=g_\rho$, with
$g_{\omega\rho\pi} = -\frac{3 g_\rho^2}{8 \pi^2 \fpi}$.
For the $\omega\rightarrow \pi\gamma$ decay,
their result is the same as ours.
However  for $\rho\rightarrow \pi\gamma$, they predict
$\Gamma(\rho\rightarrow \pi \gamma)= \Gamma(\omega\rightarrow \pi \gamma)$,
which is not in agreement with the experimental data.}
\be
g_{\omega\rho\pi} = -\frac{g_\rho g_\omega}{8 \pi^2 \fpi}.
\ee
The ellipsis stands for anomalous parity terms that are irrelevant
in our calculation.

We determine $g_\rho$ by the KSRF relation, $m_\rho^2= 2 g_\rho^2 \fpi^2$
or $g_\rho \simeq 5.85$,
and $g_\omega$ by assuming the ideal mixing,
$g_\omega=3 g_\rho\simeq 17.55$.
To see how good the anomalous parity Lagrangian is, consider
the decay widths
of $\omega\rightarrow \pi \gamma$ and
$\rho\rightarrow \pi\gamma$. Our Lagrangian predicts
\bea
\Gamma(\omega\rightarrow\pi\gamma) &=& \frac{\alpha}{24}
\left|\frac{g_{\omega\rho\pi}}{g_\rho}\right|^2\,m_\omega^3
\left(1 - \frac{\mpi^2}{m_\omega^2}\right)^3
\simeq 0.758\ \mbox{MeV},
\\
\Gamma(\rho\rightarrow\pi\gamma) &=& \frac{\alpha}{24}
\left|\frac{g_{\omega\rho\pi}}{g_\omega}\right|^2\,m_\rho^3
\left(1 - \frac{\mpi^2}{m_\rho^2}\right)^3
\simeq 0.080\ \mbox{MeV},
\eea
which are in reasonable agreement the experimental values \cite{PDG},
$\Gamma^{\rm{exp}}(\omega\rightarrow \pi\gamma)= 0.717\pm 0.050$ MeV,
$\Gamma^{\rm{exp}}(\rho^\pm\rightarrow \pi^\pm \gamma)= 0.068\pm 0.007$ MeV
and $\Gamma^{\rm{exp}}(\rho^0\rightarrow \pi^0 \gamma)=
0.121\pm 0.031$ MeV.

The characteristics of this Lagrangian can be summarized as follows:
\bitem
\item
It has chiral symmetry and other symmetries consistent with low-energy QCD.
\item It respects the vector-meson dominance, universality etc. and satisfies
the KSRF relation.
\item The chiral Lagrangian containing pions and nucleons only are recovered if
the $m_\rho$ and $m_\omega$ masses are taken to infinity.
\eitem

\begin{figure}[t]
\centerline{\epsfig{file=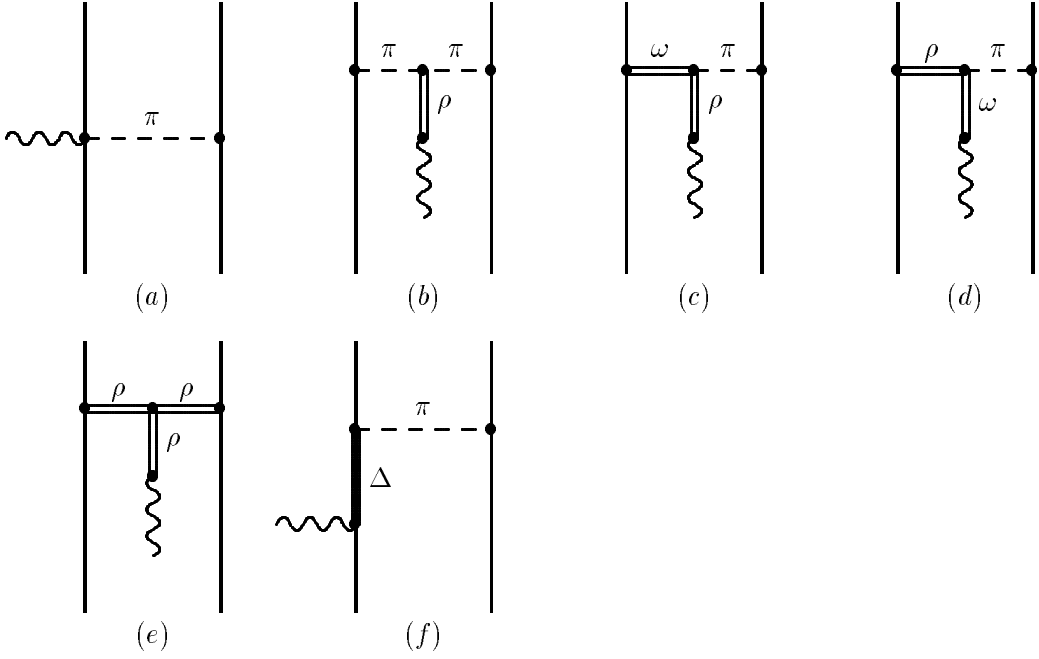}}
\caption{Graphs that contribute to the two-body vector currents.}
\label{vector:f}
\end{figure}

{}From the Lagrangian (\ref{ResL}), one sees that
only six graphs given in Fig.~\ref{vector:f}
can make
non-zero contributions to the vector current, other graphs being
either of order ${\cal O}(\mN^{-2})$ or vanishing.
They have simple physical interpretations.
Figure~\ref{vector:f}$a$ is the seagull graph with no
form factors since $a_1$-meson is absent. Figure~\ref{vector:f}$b$
is the ``pionic graph" with a form factor associated with the
$\rho$-meson dominance.
Figures~\ref{vector:f}$c$ and \ref{vector:f}$d$ stem from $\L_{an}$,
with Figure~\ref{vector:f}$d$ contributing to the baryonic current.
Figure~\ref{vector:f}$e$ is new: although it gives a non-vanishing
short-ranged Sachs moment, it cannot contribute to the ${\bm \mu}$
in two-body systems. Finally,
Figure~\ref{vector:f}$f$ is a $\Delta$-resonance contribution.
The explicit forms of the currents derived from these graphs are:
\bea
\Jem^\mu(a) &=& i\ttz \frac{\gA^2}{\fpi^2} S_1^\mu
\frac{S_2\cdot q_2}{\mpi^2-q_2^2} + (1\leftrightarrow 2),
\nonumber\\
\Jem^\mu(b) &=& -2\,i \ttz \frac{\gA^2}{\fpi^2}
\frac{m_\rho^2}{m_\rho^2-k^2}\,
\frac{S_1\cdot q_1}{\mpi^2-q_1^2}\,
\frac{S_2\cdot q_2}{\mpi^2-q_2^2}\,
\left(q^\mu - \frac{k^\mu\,k\cdot q}{m_\rho^2}\right),
\nonumber\\
\Jem^\mu(c) &=& -i \tau_2^z\, \frac{\gA}{16\pi^2 \fpi^2}\,
\frac{m_\rho^2}{m_\rho^2-k^2}\,
\frac{g_\omega^2}{m_\omega^2-q_1^2}\,
\frac{q_2\cdot S_2}{\mpi^2-q_2^2}
\, \epsilon^{\mu\nu\alpha\beta}
v_\nu k_\alpha q_\beta + (1\leftrightarrow 2),
\nonumber\\
\Jem^\mu(d) &=&
-i\,\tau_1\cdot\tau_2\,
\frac{\gA}{16\pi^2 \fpi^2}\,
\frac{m_\omega^2}{m_\omega^2-k^2}\,
\frac{g_\rho^2}{m_\rho^2-q_1^2}\,
\frac{q_2\cdot S_2}{\mpi^2-q_2^2}
\, \epsilon^{\mu\nu\alpha\beta}
v_\nu k_\alpha q_\beta + (1\leftrightarrow 2),
\nonumber\\
\Jem^\mu(e) &=& -i \ttz \frac{g_\rho^2}{2}\,
\frac{m_\rho^2}{m_\rho^2-k^2}\,\frac{1}{m_\rho^2-q_1^2}\,
\frac{1}{m_\rho^2-q_2^2}\,
\left(q^\mu - \frac{k^\mu\, k\cdot q}{m_\rho^2}\right),
\nonumber\\
\Jem^\mu(f) &=&
- \frac{\muT\,{\cal C}\, \gA}{4 \mN\,\ddm\, \fpi^2}\,
\frac{q_2\cdot S_2}{\mpi^2-q_2^2}\,
\left[
\left(\frac23 \tau_2^z - \frac{i}{3} \ttz\right)
\,q_{2 \beta} \calP_1^{\beta\alpha}
(k_\alpha\, S_1^\mu - g_\alpha^\mu\,k\cdot S_1)
\right.\nonumber \\
&&\ \ \ \ -\ \left.
\left(\frac23 \tau_2^z + \frac{i}{3} \ttz\right)
\,(S_1^\mu\,k_\alpha - g^\mu_\alpha\,k\cdot S_1)
\calP_1^{\alpha\beta} q_{2 \beta}
\right]
+ (1\leftrightarrow 2),
\label{res:VmuVector}\eea
where the projection operator $\calP_1^{\alpha\beta}$ is defined by
$$\calP_1^{\alpha\beta} = v^\alpha v^\beta - g^{\alpha\beta}
- \frac43 S_1^\alpha S_1^\beta.$$
We have dropped terms of order ${\cal O}(\mN^{-2})$ and
order $v^\mu\cdot \O(\mN^{-1})$, while keeping terms
up to $\O(\mN^{-1})$ for the space component and
up to $\O(\mN^{0})$ for the time component.
The corresponding magnetic operators then take the form
\bea
{\bm \mu}(a) &=&
e\,\frac{\gA^2 \mpi}{16\fpi^2}
\left(\frac23 \TtimesS- \TtimesT\right) \, y_1(x_\pi),
\nonumber \\
{\bm \mu}(b) &=&e\, \frac{\gA^2 \mpi}{16 \fpi^2} \left(
\frac23 \TtimesS \left[y_1(x_\pi)- 3 y_0(x_\pi)\right]
-\TtimesT\,y_1(x_\pi) \right),
\nonumber \\
{\bm \mu(c)}&=&
-\frac{e\,\gA}{64 \pi^2 \fpi^2}\, \frac{g_\omega^2}{m_\omega^2-\mpi^2}
\left( \frac13 \left(\TplusS+\TminusS\right)
\left[\mpi^3 \,y_0(x_\pi) - m_\omega^3 \,y_0(m_\omega r)\right]
\right. \nonumber \\
&&\left. \ \ \ \ \ \ \ \
+\ \left(\TplusT+\TminusT\right)\,
\left[\mpi^3 \,y_2(x_\pi) -m_\omega^3 \,y_2(m_\omega r) \right]\right),
\nonumber \\
{\bm \mu}(d)&=&
-\frac{e\,\gA}{32 \pi^2 \fpi^2}\, \frac{g_\rho^2}{m_\rho^2-\mpi^2}
\left( \frac13 \BplusS
\left[\mpi^3 \,y_0(x_\pi) - m_\rho^3 \,y_0(m_\rho r)\right]
\right. \nonumber \\
&&\left. \ \ \ \ \ \ \ \
+\ \BplusT\,\left[\mpi^3 \,y_2(x_\pi) -m_\rho^3 \,y_2(m_\omega r)
\right]\right),
\nonumber \\
{\bm \mu}(e) &=& 0,
\nonumber \\
{\bm \mu}(f)  &=&
- \frac{e\,\muT\, {\cal C}\,\gA\, \mpi^3}{36 \mN\,\ddm\,\fpi^2}
\left(\frac13\left(\TplusS + \TminusS - \TtimesS\right)
{\bar y}_0(x_\pi)
\right.
\nonumber \\
&&\left.\ \ \ \ \ \
+\ \left( \TplusT + \TminusT + \frac12 \TtimesT\right) y_2(x_\pi)
\right)
\label{murVector}\eea
where
${\bar y}_0(x)= y_0(x) - \delta({\bm x})$ and
\bea
\BplusS &\equiv& \tau_1 \cdot \tau_2\ ({\bm \sigma}_1+{\bm \sigma}_2),
\nonumber \\
\BplusT &\equiv& \tau_1\cdot\tau_2\
\left[
{\hat r} \ {\hat r}\cdot({\bm \sigma}_1+{\bm \sigma}_2)
-\frac13 ({\bm \sigma}_1+{\bm \sigma}_2)\right].
\eea
As emphasized in the main text,
the short-ranged contribution ${\bm \mu}(e)$ ({\it i.e.}, the
four-Fermi contact term) is equal to zero.



\begin{thebibliography} {100}
\bibitem{fm}
B. Frois and J.-F. Mathiot, Comments Part. Nucl. Phys. {\bf 18} (1989) 291.
\bibitem{review}
M. Rho and D.H. Wilkinson,
{\it Mesons in Nuclei} Vol. I, North-Holland, 1979;
D.O. Riska, Physica Scripta {\bf 31} (1985);
Phys. Repts. {\bf 181} (1989) 208;
J.-F. Mathiot, Phys. Repts. {\bf 173} (1989) 63;
I.S. Towner, Phys. Reports {\bf 155} (1987) 263.
\bibitem{wein79} S. Weinberg, Physica (Amsterdam) {\bf 96A} (1979) 327.
\bibitem{leut94} H. Leutwyler, Ann. Phys. (N.Y.) {\bf 235} (1994) 165.
\bibitem{wein90} S. Weinberg, \pl \ {\bf B251} (1990) 288; \np \ {\bf
B363} (1991) 3; \pl \ {\bf B295} (1992) 114.
\bibitem{vankolck} C. Ord\'o\~nez, L. Ray and U. van Kolck, \prl \ {\bf
72} (1994) 1982.
\bibitem{pmr1}
  T.-S. Park, D.-P. Min and M. Rho, Phys. Repts. {\bf 233} (1993) 341;
 T.-S. Park, I.S. Towner and K. Kubodera, \np \ {\bf A579} (1994) 381.
\bibitem{pmrnp} T.-S. Park, D.-P. Min and M. Rho,
``Radiative neutron-proton capture in effective chiral Lagrangians",
\prl, to be published.
\bibitem{riskabrown} D.O. Riska and G.E. Brown, \pl \ {\bf B38} (1972) 193.
\bibitem{villars} F. Villars, Helv. Phys. Acta {\bf 20} (1947) 476.
\bibitem{chemrho} M. Chemtob and M. Rho, \np \ {\bf A163} (1971) 1.
\bibitem{kdr} K. Kubodera, J. Delorme and M. Rho, \prl \ {\bf 40} (1978) 755.
\bibitem{HFF} E. Jenkins and A.V. Manohar,  in
{\it Proc. of the Workshop on Effective Field Theories of the Standard
Model}, Dobog\'ok\'o, Hungary, Aug. 1991, ed. U.-G. Meissner
(World Scientific, Singapore, 1992).
\bibitem{br91} G.E. Brown and M. Rho, \prl {\bf 66} (1991) 2720.
\bibitem{br95} G.E. Brown and M. Rho,
``Chiral restoration in hot and/or dense matter", Physics Repts., to be
published.
\bibitem{reviewchp} G. Ecker, hep-hp/9501357; A. Pich, hep-hp/9502366;
J. Bijnens, hep-ph/9502393; V. Bernard, N. Kaiser and U.-G.  Meissner,
hep-ph/9501384.
\bibitem{mr91} M. Rho, \prl \ {\bf 66} (1991) 1275.
\bibitem{gss}
J. Gasser, M.E. Sainio and S. \v{S}varc, \np \ {\bf B307} (1988) 779.
\bibitem{georgi}
H. Georgi, \pl \ {\bf B240} (1990) 447; ``Heavy-Quark Effective Field
Theory," in {\it Proc. of the Theoretical Advanced Study Institute 1991},
ed. R.K. Ellis, C.T. Hill and J.D. Lykken (World Scientific, Singapore, 1992)
p. 689.
\bibitem{jm91}
E. Jenkins and A. Manohar, \pl \ {\bf B255} (1991) 558; \pl \ {\bf B259}
(1991) 353; E. Jenkins, \np \ {\bf B368} (1991) 190.
\bibitem{grin} B. Grinstein, ``Light-quark, heavy-quark systems",
SSCL-Preprint-34.
\bibitem{Ecker:heat}
G. Ecker, ``Chiral invariant renormalization of the
pion--nucleon interaction", hep-ph/9402337.
\bibitem{ecker}
G. Ecker, J. Gasser, A. Pich and E. de Rafael, \np \ {\bf B321}
(1989) 311; G.Ecker, J. Gasser, H. Leutwyler, A. Pich and
E. de Rafael, \pl \ {\bf B223} (1989) 425.
\bibitem{jenkinsetal}
E. Jenkins, M. Luke, A. V. Manohar and M. Savage, \pl \ {\bf B302} (1993) 482.
\bibitem{cox} A.E. Cox, S.A.R. Wynchank and C.H. Collie,
\np \ {\bf 74} (1965) 497.
\bibitem{aust}
N. Austern, Phys. Rev. {\bf 92} (1953) 670.
\bibitem{v18} R.B. Wiringa, V.G.J. Stoks and R. Schiavilla, ``An accurate
nucleon-nucleon potential with charge independence breaking," nucl-th/9408016.
\bibitem{mallik}
S. Mallik, ``Massive states in chiral perturbation theory",
hep-ph/9410344.
\bibitem{reid} R. V. Reid, Ann. of Phys, {\bf 50} (1968) 411.
\bibitem{beane} S. Beane and U. van Kolck, \pl\ {\bf B329} (1994) 137.
\bibitem{ian} I.S. Towner, \np\ {\bf A540} (1992) 478.
\bibitem{kirsch} M. Kirchbach, D.O. Riska and K. Tsushima, \np \ {\bf
A542} (1992) 14.
\bibitem{kr} K. Kubodera and M. Rho, \prl \ {\bf 67} (1991) 3479.
\bibitem{fuji}
T. Fujiwara, T. Kugo, H. Terao and S. Uehara,
Prog. of Theo. Phys. {\bf 73} (1985) 926.
\bibitem{PDG}
Particle Data Group, \pr \ {\bf D45} (1992).

\end{thebibliography}
\end{document}